\documentclass[fleqn,usenatbib]{mnras}

\usepackage{newtxtext,newtxmath}

\usepackage[T1]{fontenc}

\DeclareRobustCommand{\VAN}[3]{#2}
\let\VANthebibliography\thebibliography
\def\thebibliography{\DeclareRobustCommand{\VAN}[3]{##3}\VANthebibliography}

\usepackage{graphicx}	
\usepackage{amsmath}	

\usepackage{tablefootnote}

\graphicspath{{./}{figures/}}

\def \msun {M$_{\odot}$}

\title[Dust polarisation in OMC-1]{OMC-1 dust polarisation in ALMA Band 7: Diagnosing grain alignment mechanisms in the vicinity of Orion Source I}

\author[K. Pattle et al.]{
Kate Pattle,$^{1, 2}$\thanks{E-mail: katherine.pattle@nuigalway.ie (KP)}
Shih-Ping Lai,$^{2, 3}$
Melvyn Wright,$^{4}$
Simon Coud\'{e},$^{5}$
Richard Plambeck,$^{4}$ \newauthor 
Thiem Hoang,$^{6}$ Ya-Wen Tang,$^{3}$ Pierre Bastien,$^{7}$ Chakali Eswaraiah$^{8}$, Ray Furuya,$^{9}$ Jihye Hwang,$^{6}$ \newauthor
Shu-ichiro Inutsuka,$^{10}$ Kee-Tae Kim,$^{6,11}$ Florian Kirchschlager,$^{12}$ Woojin Kwon,$^{13,6}$ Chang Won Lee,$^{6}$ \newauthor
Sheng-Yuan Liu,$^{3}$ Aran Lyo,$^{6}$ Nagayoshi Ohashi,$^{3}$ Mark G. Rawlings,$^{14}$ Mehrnoosh Tahani,$^{15}$ \newauthor 
Motohide Tamura,$^{16,17,18}$ Archana Soam,$^{5}$ Jia-Wei Wang,$^{3}$ and Derek Ward-Thompson$^{19}$
\\
$^{1}$National University of Ireland Galway, University Road, Galway H91 TK33, Ireland\\
$^{2}$Center for Astronomy, National Tsing Hua University, No. 101, Sec. 2 Guangfu Road, Hsinchu 30013, Taiwan\\
$^{3}$Academia Sinica Institute of Astronomy and Astrophysics, No. 1, Sec. 4., Roosevelt Road, Taipei 10617, Taiwan \\
$^{4}$Radio Astronomy Lab, University of California, 501 Campbell Hall, Berkeley CA 94720-3441, USA\\
$^{5}$SOFIA Science Center, Universities Space Research Association, NASA Ames Research Center, Moffett Field, California 94035, USA\\
$^{6}$ Korea Astronomy and Space Science Institute, 776 Daedeokdae-ro, Yuseong-gu, Daejeon 34055, Republic of Korea \\
$^{7}$ Centre de recherche en astrophysique du Qu\'{e}bec \& d\'{e}partement de physique, Universit\'{e} de Montr\'{e}al, C.P. 6128 Succ. Centre-ville, \\ Montr\'{e}al, QC, H3C 3J7, Canada \\
$^{8}$ CAS Key Laboratory of FAST, National Astronomical Observatories, Chinese Academy of Sciences, People's Republic of China \\
$^{9}$ Institute of Liberal Arts and Sciences, Tokushima University, Minami Jousanajima-machi 1-1, Tokushima 770-8502, Japan \\
$^{10}$ Department of Physics, Graduate School of Science, Nagoya University, Furo-cho, Chikusa-ku, Nagoya 464-8602, Japan \\
$^{11}$ University of Science and Technology, Korea, 217 Gajeong-ro, Yuseong-gu, Daejeon 34113, Republic of Korea \\
$^{12}$ Department of Physics and Astronomy, University College London, Gower Street, London WC1E 6BT, UK \\
$^{13}$ Department of Earth Science Education, Seoul National University, 1 Gwanak-ro, Gwanak-gu, Seoul 08826, Republic of Korea \\
$^{14}$ East Asian Observatory, 660 N. A'oh\={o}k\={u} Place, University Park, Hilo, HI 96720, USA \\
$^{15}$ Dominion Radio Astrophysical Observatory, Herzberg Astronomy and Astrophysics Research Centre, National Research Council Canada, \\
P. O. Box 248, Penticton, BC V2A 6J9 Canada \\
$^{16}$ National Astronomical Observatory of Japan, National Institutes of Natural Sciences, Osawa, Mitaka, Tokyo 181-8588, Japan \\
$^{17}$ Department of Astronomy, Graduate School of Science, The University of Tokyo, 7-3-1 Hongo, Bunkyo-ku, Tokyo 113-0033, Japan \\
$^{18}$ Astrobiology Center, National Institutes of Natural Sciences, 2-21-1 Osawa, Mitaka, Tokyo 181-8588, Japan \\
$^{19}$ Jeremiah Horrocks Institute, University of Central Lancashire, Preston PR1 2HE, UK 
}

\date{Accepted XXX. Received YYY; in original form ZZZ}

\pubyear{2021}

\begin{document}
\label{firstpage}
\pagerange{\pageref{firstpage}--\pageref{lastpage}}
\maketitle

\begin{abstract}

We present ALMA Band 7 polarisation observations of the OMC-1 region of the Orion molecular cloud.  We find that the polarisation pattern observed in the region is {likely to have been} significantly altered by the radiation field of the $>10^{4}$\,L$_{\odot}$ high-mass protostar Orion Source I.  {In the protostar's optically thick disc, polarisation is likely to arise from dust self-scattering.  In material to the south of Source I -- previously identified as a region of `anomalous' polarisation emission -- we observe a polarisation geometry concentric around Source I.  We demonstrate that Source I's extreme luminosity may be sufficient to make the radiative precession timescale shorter than the Larmor timescale for moderately large grains ($> 0.005-0.1\,\mu$m), causing them to precess around the radiation anisotropy vector (k-RATs) rather than the magnetic field direction (B-RATs).  This requires relatively unobscured emission from Source I, supporting the hypothesis that emission in this region arises from the cavity wall of the Source I outflow.  This is one of the first times that evidence for k-RAT alignment has been found outside of a protostellar disc or AGB star envelope.  Alternatively, the grains may remain aligned by B-RATs and trace gas infall onto the Main Ridge.  Elsewhere, we largely find the magnetic field geometry to be radial around the BN/KL explosion centre, consistent with previous observations.  However, in the Main Ridge, the magnetic field geometry appears to remain consistent with the larger-scale magnetic field, perhaps indicative of the ability of the dense Ridge to resist disruption by the BN/KL explosion.}

\end{abstract}


\section{Introduction} \label{sec:intro}

The role of magnetic fields in star formation, and particularly in high-mass star formation, remains poorly constrained.  Until recently, there was a lack of observational evidence for the magnetic field morphology in the high-density interstellar medium (ISM).  Dust emission polarimetry is a long-standing means of inferring ISM magnetic field properties \citep{davis1951}; however, observations have in the past been strongly surface-brightness-limited.  The polarimetric capabilities of the new generation of submillimetre telescopes, including the Atacama Large Millimeter/submillimeter Array (ALMA), have made magnetic fields in dense, star-forming gas newly accessible \citep[e.g.][]{cortes2016,kwon2019}.  However the density regimes now observable have brought with them complications in interpretation of polarisation observations, as the mechanisms by which dust grains can gain a preferential alignment proliferate at high densities \citep{davis1951,gold1952,lazarian2007a,lazarian2007b,kataoka2015,hoang2018,kataoka2019}.

\begin{figure*}
    \centering
    \includegraphics[width=\textwidth]{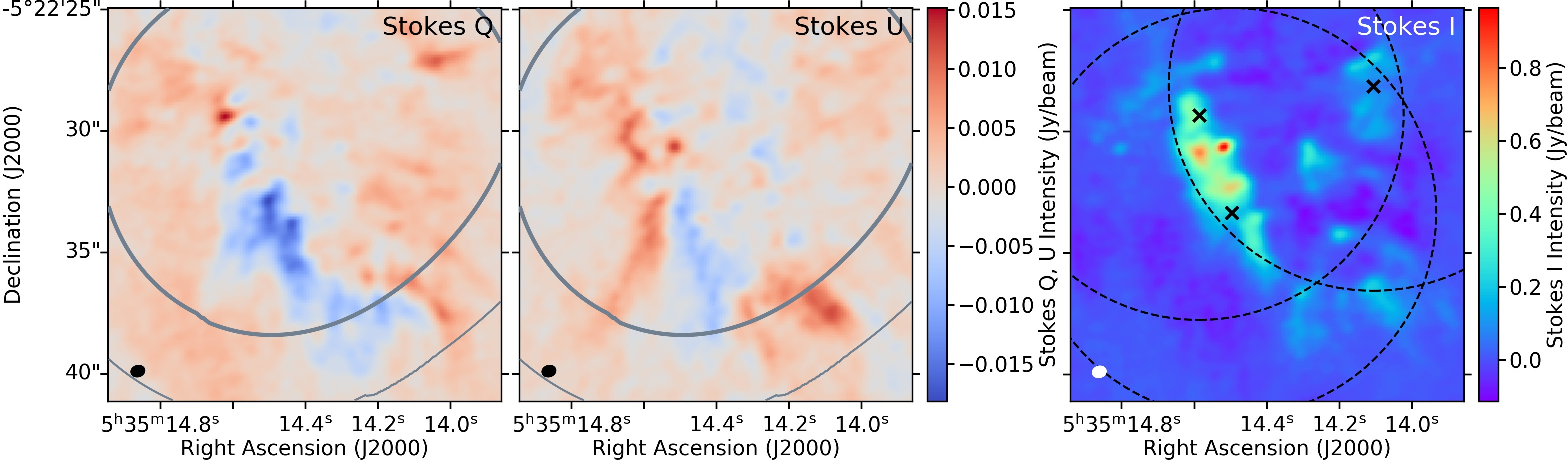}
    \caption{Mosaicked Stokes $Q$ (left), $U$ (centre) and $I$ (right) maps {ALMA Band 7 ({340.6\,GHz, 881$\mu$m})} maps of OMC-1.  The synthesised beam is shown in the lower left-hand corner of each plot.  {The 50\% (HPBW) and 90\% gain contours of the combined primary beam are shown as thin and thick grey lines respectively in the left and centre panels.  The centres of our three fields are marked with crosses on the right panel, and their primary beam HPBWs (16.8$^{\prime\prime}$) are marked with dashed lines.}}
    \label{fig:QU}
\end{figure*}

Most observations of submillimetre dust polarisation at very high densities have been in protostellar discs \citep[e.g.][and refs. therein]{hull2019}.  However, sites of high-mass star formation are another important high-density ISM environment.  In this work, we present ALMA Band 7 {(340.6\,GHz; 881\,$\mu$m)} observations of the OMC-1 region, at the centre of the Orion Molecular Cloud, a nearby site of high-mass star formation \citep[e.g.][]{bally2008}.

The OMC-1 region, at the centre of the well-studied `integral filament' in the Orion A molecular cloud, is located at a distance of $388\pm 5$\,pc \citep{kounkel2017}, and consists of two dense clumps - the northern Becklin-Neugebauer-Kleinmann-Low (BN/KL) clump \citep{becklin1967, kleinmann1967}, and the southern Orion S clump \citep{batria1983, haschick1989}.  In this paper we focus on the centre of the BN/KL clump, an active site of star formation which hosts an extremely powerful wide-angle explosive molecular outflow, with multiple ejecta known as the `bullets of Orion' \citep{kwan1976, allen1993}.  The young stars BN, Source I, and $x$, located in the core of the BN/KL clump, have proper motions consistent with their having been co-located $\sim 500$ years ago, leading to the suggestion that the BN/KL outflow is the result of a dynamical interaction between these sources \citep{gomez2005}. The dynamic age of the BN/KL outflow is also $\sim 500$\,yr \citep{zapata2009}, and the kinetic energy released by the interaction is comparable to the energy in the outflow \citep{kwan1976,gomez2005}, supporting this interpretation.  An alternative explanation for the BN/KL outflow is a protostellar merger \citep{bally2005}.  Debate over what combination of interaction, decay and merger produced the BN/KL outflow continues \citep[e.g.][]{luhman2017,farais2018}; however, the approximate age, high energy and impulsive nature of the outflow are well-established. 

Source I drives a separate, slower, bipolar outflow along an axis perpendicular to its direction of motion \citep{plambeck2009}.  BN and Source I appear to be recoiling from a common centre \citep{rodriguez2005, luhman2017}; nonetheless, the outflow from Source I is symmetric about an axis approximately perpendicular to the direction of motion, despite the significant ram pressure on the source as it ploughs through its surroundings at $\sim 12$\,km\,s$^{-1}$ \citep{rodriguez2005}.  This discrepancy could be ascribed to the outflow being shaped by a strong magnetic field.  \citet{hirota2020} recently observed a highly uniform polarisation structure in SiO emission associated with the outflow, suggesting a field strength of $\sim 30$\,mG, strong enough to prevent distortion of the outflow by ram pressure.

OMC-1 has, on large scales, an hourglass magnetic field \citep{schleuning1998,houde2004,wardthompson2017,pattle2017}.  The clump which we observe, in the centre of the OMC-1 region, has an approximately linear magnetic field across it (\citealt{chrysostomou1994}, \citealt{simpson2006}, \citealt{pattle2017}), with an orientation $-64^{\circ}.2\pm6^{\circ}.5$ E of N \citep{wardthompson2017}.  Estimates of the plane-of-sky field strength in OMC-1 range from $\sim 1-10$\,mG \citep{hildebrand2009,houde2009,pattle2017}, all of which indicate a strong magnetic field. 

Complete depolarisation is observed on the position of BN/KL in single-dish observations \citep{schleuning1998,houde2004,pattle2017}.  This depolarisation, over a single telescope beam, results from an approximately elliptical polarisation pattern in the dense centre of OMC-1, as observed using BIMA \citep{rao1998} and the SMA \citep{tang2010}.  In regions where dust grains are aligned with their major axes perpendicular to the magnetic field direction, polarisation vectors can be rotated by 90 degrees to trace the plane-of-sky  magnetic field \citep{davis1951, andersson2015}.  Thus \citet{tang2010}, observing the 870$\mu$m dust continuum with the SMA, inferred that the magnetic field in the region is radial, and centred on the outflow. From this they suggested two hypotheses: (1) a toroidal field in a magnetised, differentially rotating `pseudo-disk' in the centre of OMC-1, or
(2) the magnetic field is being dragged into a radially-symmetric morphology by the explosive outflow.

{OMC-1 has recently been observed in ALMA Bands 3 (3.1 mm) and 6 (1.3mm) by \citet{cortes2020}.  They find a magnetic field of strength $9.4\pm1.8\,$mG with a `quasi-radial' configuration centred on the position of the BN/KL explosion.  Their energetics analysis suggests both that the magnetic field is well-coupled to the gas and that the energy in the field is much less than that in the BN/KL explosive outflow, favouring the second of the \citet{tang2010} hypotheses.  In this paper we present ALMA Band 7 (0.88mm) observations of Source I and its surroundings, a subset of the area observed by \citet{cortes2020}.  We investigate whether a single model, or a single grain alignment mechanism, is sufficient to explain the complex polarisation morphology observed in the vicinity of Source I.}

In recent years, a number of different grain alignment mechanisms have been suggested to explain the polarisation properties of dust emission in different environments in the very high-density ISM.  As well as the traditional interpretation of polarised dust emission as tracing the plane-of-sky magnetic field direction, an effect usually ascribed to Radiative Alignment Torques (B-RATs; \citealt{lazarian2007a, andersson2015}), alternative mechanisms include supersonic mechanical grain alignment (the Gold effect; \citealt{gold1952}), Mechanical Alignment Torques (MATs; \citealt{lazarian2007b,hoang2016}), {a variation on Radiative Alignment Torques in which grains precess around the radiation anisotropy vector rather than the magnetic field direction (k-RATs; \citealt{lazarian2007a, tazaki2017})}, and dust self-scattering \citep{kataoka2015}.  In this work we will discuss the polarisation pattern of the {OMC-1} region in the context of these various mechanisms by which polarised emission can arise.

The structure of this paper is as follows: in Section~\ref{sec:data} we describe our observations.  In Section~\ref{sec:results} we interpret the polarisation distributions seen across the region.  Our conclusions are summarised in Section~\ref{sec:summary}.

\section{Data} \label{sec:data}

\begin{figure*}
    \centering
    \includegraphics[width=\textwidth]{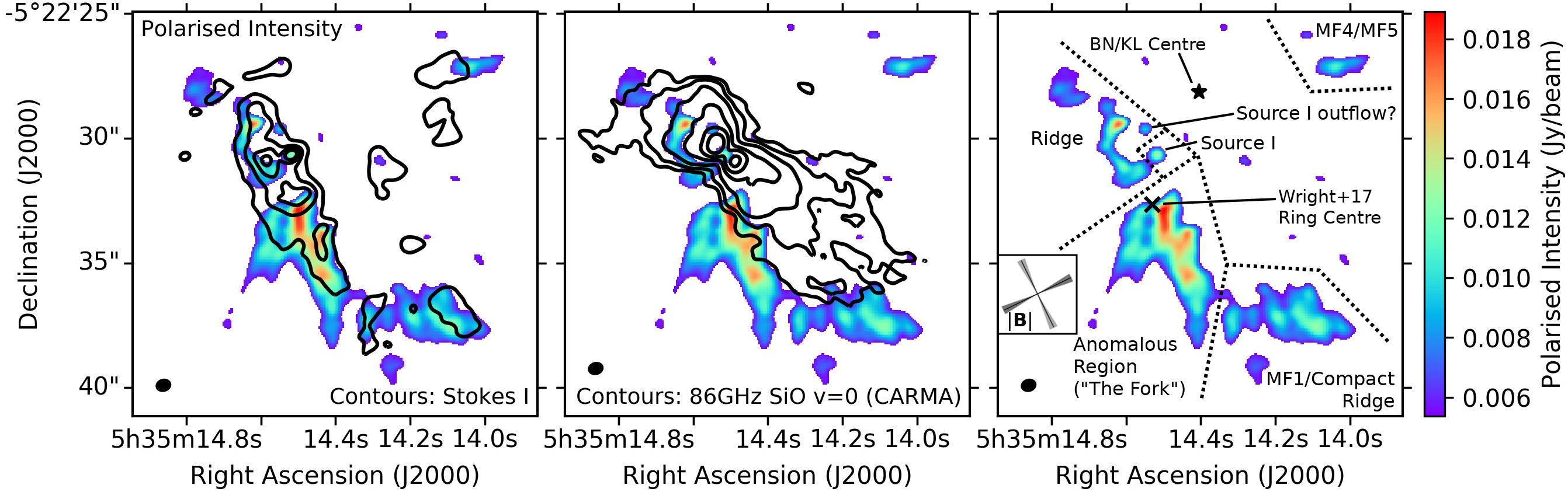}
    \caption{Mosaicked polarised intensity map of OMC 1.  Left: Polarised intensity overlaid with contours of total intensity (Stokes $I$).  Contour levels are 0.1, 0.3, 0.5 and 0.7 Jy/beam.  Note that the Stokes $I$ map is more dynamic-range-limited than is the polarised intensity map.  Centre: Polarised intensity overlaid with contours of CARMA 86\,GHz SiO $v=0$ emission averaged over the velocity range $-10$ to $+20$\,km\,s$^{-1}$, tracing the Source I outflow \citep{plambeck2009}.  Contour levels are 0.05, 0.1, 0.2, 0.5, 1.0, and 2.0 Jy/CARMA beam. Right: An illustration of the division of regions in OMC-1 used in this work.  The centre of the BN/KL explosion \citep{rodriguez2017} is marked with a black star and the centre of the \citet{wright2017} ring feature is marked with a black cross.  The mean magnetic field direction observed on large scales ({$115.8\pm6.5^{\circ}$} E of N) is shown {in the inset box} as a vector {with a dark grey wedge indicating uncertainty}, and its associated polarisation direction ($25.8\pm 6.5^{\circ}$ E of N) is shown as a vector {with a light grey wedge}.  In each panel, the synthesised beam is shown in the lower left-hand corner.}
    \label{fig:PI}
\end{figure*}

We observed three overlapping fields in OMC-1 in ALMA Band 7 polarised light.  These observations were taken in ALMA Cycle 6 on 2019 April 9.  We observed one track on each source, for a total of 1.5 hours of observing time, in array configuration C43-3.  The data have project code 2018.1.01162.S.  The three fields were centred on the Orion Hot Core, with R.A. (J2000) $=\,05^{h}35^{m}14^{s}.59$, Dec. (J2000)  $=\,05^{\circ}22^{\prime}29.^{\prime\prime}5$; SMA 1, with R.A. (J2000) $=\,05^{h}35^{m}14^{s}.50$, Dec. (J2000) $\,=05^{\circ}22^{\prime}33.^{\prime\prime}5$; and the Northwest Clump, with R.A. (J2000) $\,=05^{h}35^{m}14^{s}.11$, Dec. (J2000) $=\,05^{\circ}22^{\prime}28.^{\prime\prime}3$.  

{The central sky frequency of the observations of each field is 340.7\,GHz (881\,$\mu$m).  We observed four spectral windows, three continuum (spw0-2) and one line (spw3).  The three continuum windows have central frequencies of 333.8\,GHz (899\,$\mu$m), 335.6\,GHz (894\,$\mu$m) and 347.6\,GHz (863\,$\mu$m), and effective bandwidths of 2000\,MHz.  The spectral line window has a central frequency of 345.8\,GHz (868\,$\mu$m), with an effective bandwidth of 1875\,MHz, and a channel width of 3.906\,MHz.  The spectral line window is centred on the $^{12}$CO $J=3\to2$ line, as discussed below.}

The data were calibrated and imaged using CASA version 5.4.0, {using imaging scripts supplied} by the Observatory.  The phase calibrator was J0529-0519 and the polarisation calibrator was J0522-3627.  The $tclean$ parameters used for imaging were Briggs weighting, with a robust parameter of 0.5, and a cell size of 0.060$^{\prime\prime}$.  The output maps have a restoring beam size of $0.54^{\prime\prime}\times0.43^{\prime\prime}$, oriented $-74^{\circ}$ E of N, and a maximum recovered size scale of 4.7\arcsec.  Integrated Stokes $Q$, $U$ and $I$ maps were produced {using the data from spectral windows spw0-2}; note that we do not consider circular polarisation (Stokes $V$) in this work.

{Emission from OMC-1 includes contributions from a plethora of spectral lines \citep[e.g.][]{pagani2017}.  Fortunately for our purposes, most of these lines contribute little polarised signal, and so our Stokes $Q$ and $U$ maps made from spw0-2 are dominated by continuum emission.  The most significant spectral line in our set of spectral windows is the $^{12}$CO $J=3\to2$ line at 345.8\,GHz, {on which spw3 is centred}.  CO emission in outflows can be polarised up to a maximum of $\sim 3$\% by the Goldreich-Kylafis effect (GK effect; \citealt{goldreich1982,ching2016}), which has previously been observed in OMC-1 \citep{houde2013}.  Data from the spectral window dedicated to imaging this line (spw3) is therefore not considered in this work.  We will analyse CO line polarisation in OMC-1 in a future work.}

We linearly mosaicked {our continuum Stokes $Q$, $U$ and $I$} maps using the Miriad task $linmos$ 
{which performs a primary beam weighted mosaic of the three pointing centers to minimise the RMS error. We used the taper option to obtain an approximately uniform noise across the image \citep{sault1996}.
The individual images of the three pointings are in good agreement
with the resulting mosaic. The effective
combined primary beam weighting is shown in
Figure~\ref{fig:QU}. The three pointings are
within the FWHM of the primary beam, which minimises any effects of primary beam off-axis polarisation.
}
The mosaicked $Q$, $U$ and $I$ maps are shown in Figure~\ref{fig:QU}.  Our RMS noise in the mosaicked Stokes $Q$ and $U$ maps is 0.62 mJy/beam.  For the purpose of the analysis in Section~\ref{sec:results} below, we regridded the Stokes $Q$, $U$ and $I$ maps to a 0.25\arcsec\ pixel grid, approximately Nyquist-sampled on the major axis of the beam.

Polarised intensity is calculated as
\begin{equation}
    PI = \sqrt{Q^{2}+U^{2}},
\end{equation}
and is shown in Figure~\ref{fig:PI}.  Polarised intensity is thresholded at 6 mJy/beam.

\begin{figure*}
    \centering
    \includegraphics[width=\textwidth]{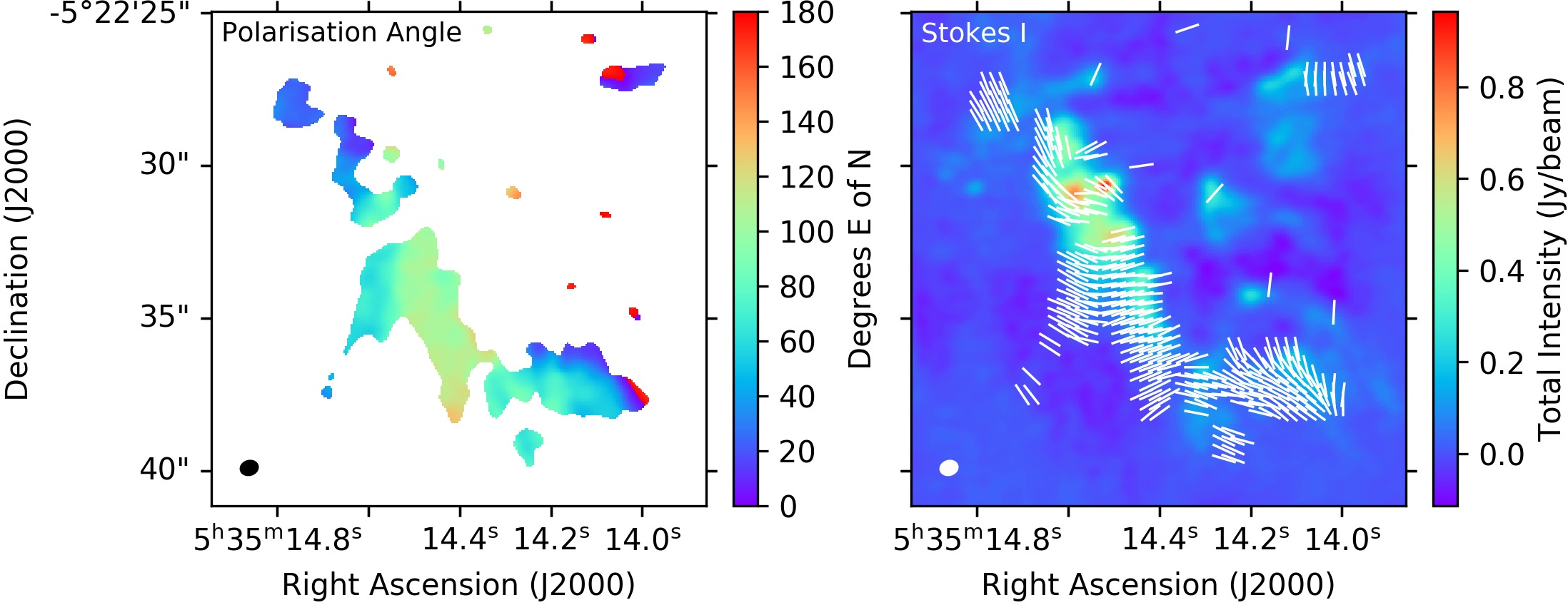}
    \caption{{Polarisation angle measurements in OMC-1 determined from mosaicked $Q$ and $U$ maps.}  Left: polarisation angle, in degrees E of N, mapped to the range $0\leq \theta<180$ deg.  Right: Polarisation vectors, plotted on total intensity (Stokes $I$) map.  In both panels, the synthesised beam is shown in the lower left-hand corner.}
    \label{fig:pa}
\end{figure*}

Polarisation angle is given by
\begin{equation}
    \theta = \frac{1}{2}\arctan\left(\frac{U}{Q}\right).
\end{equation}
Polarisation angles are shown in Figure~\ref{fig:pa}.  We note that while polarisation angle segments are referred to as vectors for convenience, they are not true vectors due to the $\pm 180^{\circ}$ ambiguity in polarisation direction.  {The polarisation geometry which we observe is in qualitative agreement with recent Band 3 and and Band 6 results at comparable resolution \citep{cortes2020}.}

{When RMS uncertainties on Stokes $Q$ and $U$ emission are equal, uncertainty on polarisation angle is given in radians by}
\begin{equation}
    \delta\theta = \frac{1}{2}\,\frac{\delta Q}{PI}.
\end{equation}
{For $\delta Q = \delta U = 0.62$\,mJy/beam and $PI\geq 6$\,mJy/beam, $\delta \theta \leq 3.0^{\circ}$.}

We do not calculate polarisation fractions in this work, as the Stokes $I$ map is much more dynamic-range-limited than are the Stokes $Q$ and $U$ maps, as shown in Figure~\ref{fig:PI}, and so any calculation of polarisation fraction is likely to produce artificially large values, particularly in low-surface-brightness regions.

\section{Results and Discussion} \label{sec:results}

\begin{table}
    \centering
    \begin{tabular}{c cc c}
    \hline
     & Mean & Median & Independent \\
    Sub-region & (deg) & (deg) & Beams \\
    \hline
    Source I & $53.5\pm8.6$ & $52.0\pm5.4$ & $< 2$ \\
    Anomalous Region/Fork & $93.7\pm19.8$ & $-88.0\pm8.7$ & 55 \\
    Main Ridge & $41.1\pm25.7$ & $29.4\pm 9.6$ & 23 \\
    MF4/MF5\tablefootnote{`MF' refers to methyl formate peaks identified by \citet{favre2011}.} & $6.7\pm8.3$ & $7.7\pm7.1$ & 6 \\
    Compact Ridge/MF1\footnotemark[1] & $57.2\pm 28.9$ & $58.9\pm15.1$ & 40 \\
    \hline
    \end{tabular}
    \caption{Sub-region polarisation angle statistics.  {Angles are given in degrees E of N.  Number of independent beams is calculated as the ratio of observed area to beam area.}}
    \label{tab:stats}
\end{table}

\begin{figure*}
    \centering
    \includegraphics[width=\textwidth]{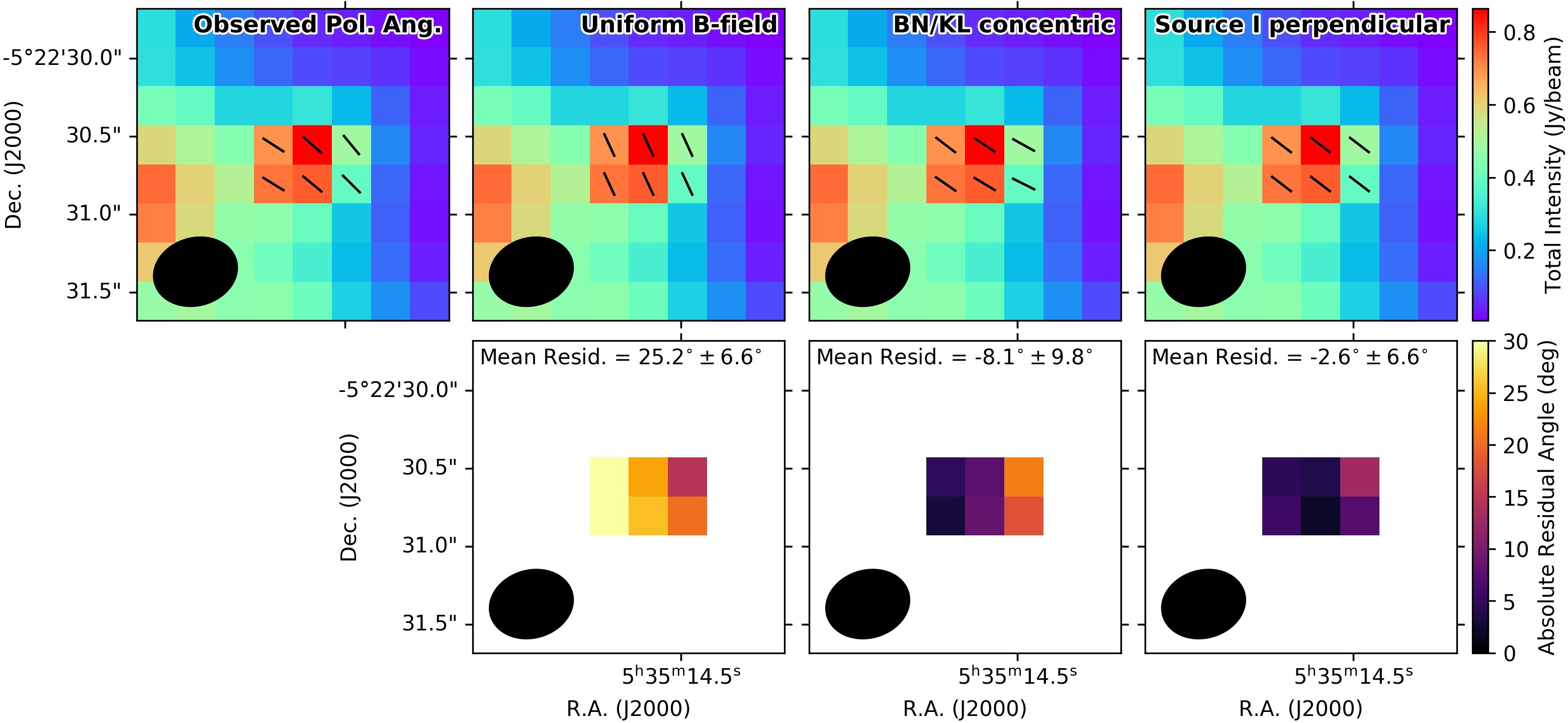}
    \caption{Comparison of models in Source I.  Top row shows observed and model polarisation geometries, plotted on Stokes $I$ emission, bottom row shows absolute difference in angle between data and models.  Far left: Observed polarisation vectors.  Centre left: polarisation vectors aligned $26^{\circ}$ E of N, perpendicular to the the large-scale magnetic field direction (hypothesised alignment mechanism: B-RATs).  Centre {right}: polarisation vectors concentric around the BN/KL explosion centre (hypothesised alignment mechanism: {B-RATs/v-MATs}).  {Far right}: polarisation vectors perpendicular to the major axis of the Source I disc (parallel to minor axis; polarisation hypothesised to arise from dust self-scattering).  {In the bottom row, the colour table saturates at a difference in angle of $30^{\circ}$, to emphasise the differences between the models.}  All maps are shown on 0.25\arcsec\ (approximately Nyquist-sampled) pixels.  The synthesised beam size is shown in the lower left-hand corner of each plot.}
    \label{fig:compare_SrcI}
\end{figure*}

\begin{figure}
    \centering
    \includegraphics[width=0.47\textwidth]{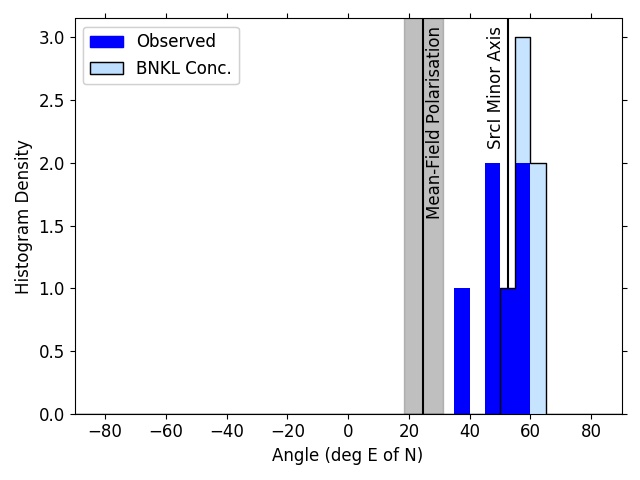}
    \caption{Histogram of polarisation angles in Source I (blue), compared with model polarisation vectors concentric around the centre of the BN/KL explosion (light blue, solid outline), {measured on 0.25\arcsec\ (approximately Nyquist-sampled) pixels}.  The polarisation angle associated with the mean 116-degree magnetic field direction is marked, as is the angle of the minor axis of the Source I disc.}
    \label{fig:histogram_SrcI}
\end{figure}

Throughout most of the ISM, polarised dust emission can be reliably assumed to arise from dust grains aligned with their major axis perpendicular to the magnetic field direction; {the leading theory for explaining which is Radiative Alignment Torques \citep[B-RATs; ][]{lazarian2007a}}.  However, at extremely high densities, such as those found in the centre of OMC-1, this assumption starts to break down.  In extreme environments such as these, possible cause of polarised dust emission include: (1) {Alignment by B-RATs, the standard mechanism throughout the ISM}.  (2) Dust self-scattering, an effect seen in protoplanetary discs, in which polarisation arises from Rayleigh scattering from large dust grains \citep{kataoka2015}.  (3) Gold alignment: mechanical alignment of dust grains in a supersonic gas flow, with grain major axes parallel to the flow direction \citep{gold1952}.  (4) Mechanical alignment torques (MATs; \citealt{lazarian2007b}), with grain major axes aligned (a) perpendicular to magnetic field direction \citep[B-MATs,][]{hoang2018}, (b) with grain major axes perpendicular to gas/dust drift direction \citep[{v-MATs,}][]{hoang2018,kataoka2019}, (5) alignment by Radiative Alignment Torques such that grain major axes are perpendicular to local radiation gradient \citep[k-RATs; ][]{lazarian2007a,tazaki2017}.  We consider these alternatives when interpreting the polarised emission from dust in OMC-1, {introducing each mechanism in detail as it arises}.

When considering possible causes of polarised emission in OMC-1, we divided the regions into the following sub-regions: (1) Source I, (2) the Anomalous Region, also referred to as the Fork, (3) the Ridge, (4) MF4/MF5, (5) the Compact Ridge, also referred to as MF1.  These sub-regions are labelled on Figure~\ref{fig:PI}.  The mean and median polarisation angles, as well as the number of independent measurements, in each region are listed in Table~\ref{tab:stats}.  We discuss each region in turn below.

Throughout the following discussion we principally compare the polarisation geometry in each region to three models: polarisation arising from grains aligned (1) with their major axes perpendicular to the large-scale 116$^{\circ}$ E of N magnetic field (polarisation angle $26^{\circ}$ E of N) {observed on larger scales in the region,} (2) such that their major axes trace concentric circles around the centre of the BN/KL explosion, and (3) such that their major axes trace concentric circles around Source I.  {In case (1) we expect to find some variation from the large-scale mean field direction, which is measured at resolutions $\gtrsim 10$\arcsec\ \citep{houde2004,wardthompson2017}, but use the mean direction as a simple model for purposes of comparison.}
We also consider the dust self-scattering model for Source I, as discussed below.

\subsection{Source I}
\label{sec:srcI}

Source I is a well-studied highly luminous high-mass protostar with a collimated SiO outflow \citep[e.g.][]{hirota2015}, as shown in Figure~\ref{fig:PI}.  Its mass has been a matter of discussion, with its velocity, approximately half that of the 10\,\msun\ B star BN, from which Source I appears to be recoiling, suggesting a mass $\sim 20$\,\msun\ \citep{rodriguez2005}, while high angular resolution observations of the rotation curves of H$_2$O and salt lines imply a central mass of 15$M_{\odot}$ \citep{ginsburg2018}, {as does recent analysis of the combined proper motions of Sources BN, I and $x$ \citep{bally2020}}. However, rotation curves of emission lines from the base of the bipolar outflow suggest a mass in the range $\sim 5-8$\,\msun\ \citep{kim2008,matthews2010,plambeck2016,Hirota2017,kim2019}.  Moreover, the high velocity of the source with respect to the surrounding dense medium is belied by the symmetry of its outflow, which would be expected to be significantly bowed by ram pressure \citep{hirota2020}.  These apparent contradictions have recently been reconciled by polarisation observations of the SiO emission associated with the outflow, suggesting a magnetic field sufficiently strong to shape the bipolar outflow and to cause sub-Keplerian gas dynamics at the base of the outflow, leading to the mass underestimate \citep{hirota2020}.

The polarisation geometry of Source I is shown in Figures~\ref{fig:compare_SrcI} and \ref{fig:histogram_SrcI}.  Source I is {effectively unresolved in our observations.  We associate six Nyquist-sampled pixels (less than two independent measurements) with the source.  The disc itself, with physical size $\sim 100$\,au \citep{ginsburg2018} and a major axis of 0.23$^{\prime\prime}$ at 340\,GHz \citep{wright2020} is smaller than the beam.}  The mean polarisation angle of Source I  {over those six pixels} is $53^{\circ}.5\pm8^{\circ}.6$, and the median is $52^{\circ}.0\pm5^{\circ}.4$, inconsistent with the polarisation having arisen from a $116^{\circ}\pm 6.5^{\circ}$ magnetic field, but consistent with being parallel to the minor axis of Source I ($53^{\circ}$; \citealt{ginsburg2018}).  The polarisation {direction which we observe} is also broadly consistent with that expected for a polarisation pattern concentric around BN/KL.  The three model geometries which we consider for Source I are shown in Figure~\ref{fig:compare_SrcI}.

\citet{kataoka2015} introduced the dust self-scattering mechanism for producing polarised emission in protostellar discs, wherein polarisation arising at a given wavelength arises from Rayleigh scattering from dust grains with sizes comparable to that wavelength.  This mechanism can produce a polarisation pattern concentric around the protostellar position, or aligned with the disc minor axis, consistent with what we see in Source I.  The conditions for polarisation arising from dust self-scattering to produce uniform polarisation aligned with the disc minor axis, as seen in Source I, are given by \citet{sadavoy2019} as an inclined (i$>$60$^{\circ}$) disc with optically thick dust emission.

\citet{wright2020} find a spectral index $\sim$2 along the disc mid-plane consistent with optically thick dust emission. This spectral index increases to $\sim$3 at the disc edges, suggesting that dust emission is optically thin in the periphery of the disc.
\citet{wright2020}, observing at 340\,GHz, set a lower limit to the disc inclination of $79^{\circ}\pm1^{\circ}$ (with major and minor axes $99\times19$ au) at a disc brightness temperature contour of 400\,K, and $74^{\circ}\pm1^{\circ}$ ($239\times45$ au) at a disc brightness temperature contour of 25K, but note that the observed geometry suggests that the inclination is closer to 90$^{\circ}$.  Similarly, \citet{matthews2010} measure an inclination $\sim85^{\circ}$ from SiO masers close to the disc.
Source I thus meets the conditions for our observed {average polarisation direction} to arise from dust self-scattering.  While we note that the polarisation geometry is also consistent with being concentric around BN/KL, dust self-scattering is an established mechanism for producing polarisation in protostellar discs, the necessary conditions for which are matched in Source I, and so we consider it to be the probable source of the dust polarisation which we observe.  This implies the existence of significant dust growth and coagulation within the Source I disc.  While grain growth is expected in protostellar discs \citep[e.g.][]{kwon2009}, for dust self-scattering to be observed, there must be a significant population of spherical dust grains with size $\sim \lambda/2\pi$ \citep{kataoka2015}, or non-spherical grains with sizes $\gtrsim \lambda/2\pi$ \citep{kirchschlager2020}.  At 870$\mu$m, this implies the existence of a population of dust grains with sizes $\sim 140\,\mu$m or larger in the Source I disc.

We note that \citet{hirota2020} found an upper limit continuum polarisation fraction of 1\% in the Source I disc at 96 GHz (3.1\,mm).  These observations, with a synthesised beam size of 0.05\arcsec, did resolve the Source I disc, and so if scattering were important, some polarisation signal would be expected.  If scattering is significant at 345\,GHz but negligible at 96\,GHz, this puts strong constraints on the grain size distribution in the Source I disc, with a significant population of grains with sizes $\sim 140\,\mu$m, but a cut-off in grain size at $<\,500\,\mu$m.  {Resolved} polarisation observations of Source I at 345\,GHz are required to confirm the dust self-scattering hypothesis, and to better constrain the grain size distribution in the disc.

\subsection{Anomalous region/Fork}
\label{sec:anom}

\begin{table*}
    \centering
    \begin{tabular}{p{3cm} p{2.3cm} p{2cm} p{2.3cm} p{1.1cm} p{1.8cm} p{2.3cm}} 
    \hline
         & B-RATs & Dust \newline self-scattering & Gold \newline alignment & v-MATs \newline (BN/KL) & v-MATs \newline (Src I outflow) & k-RATs \newline (Src I) \\
    \hline
    Predicted/observed \newline geometries consistent? & {N/A} \newline  & Yes & No & Marginal & Uncertain & Yes \\
    Polarisation arises from \newline ambient medium or \newline Src I outflow cavity walls? & Either & OCW & AM: BN/KL-driven \newline OCW: Src I-driven & AM & OCW & {OCW strongly \newline favoured} \\
    Required conditions & $\tau_{Lar}$ shortest & $>$\,100\,$\mu$m grains \newline in Src I outflow & $\Delta v > c_{s}$ & \multicolumn{2}{l}{$\tau_{mech}$ shortest; $\Delta v < c_{s}$} & $\tau_{rad,p}$ shortest \\
    Conditions met? & For small grains \newline ($a$\,$\lesssim$\,0.005--0.1$\,\mu$m) & Unlikely & Ruled out \newline geometrically  & \multicolumn{2}{l}{No; requires $\omega\ll\omega_{th}$} & For large grains \newline ($a$\,$\gtrsim$\,0.005--0.1\,$\mu$m) \\
    \hline
    \end{tabular}
    \caption{A summary of potential mechanisms for producing polarised dust emission in the {western arm of the Fork}.  `AM' {(ambient medium)} refers to polarisation in the Fork arising from the ambient medium of OMC-1; `OCW' {(outflow cavity walls)} to polarisation arising from the cavity walls of the Source I outflow.}
    \label{tab:fork}
\end{table*}

\begin{figure*}
    \centering
    \includegraphics[width=\textwidth]{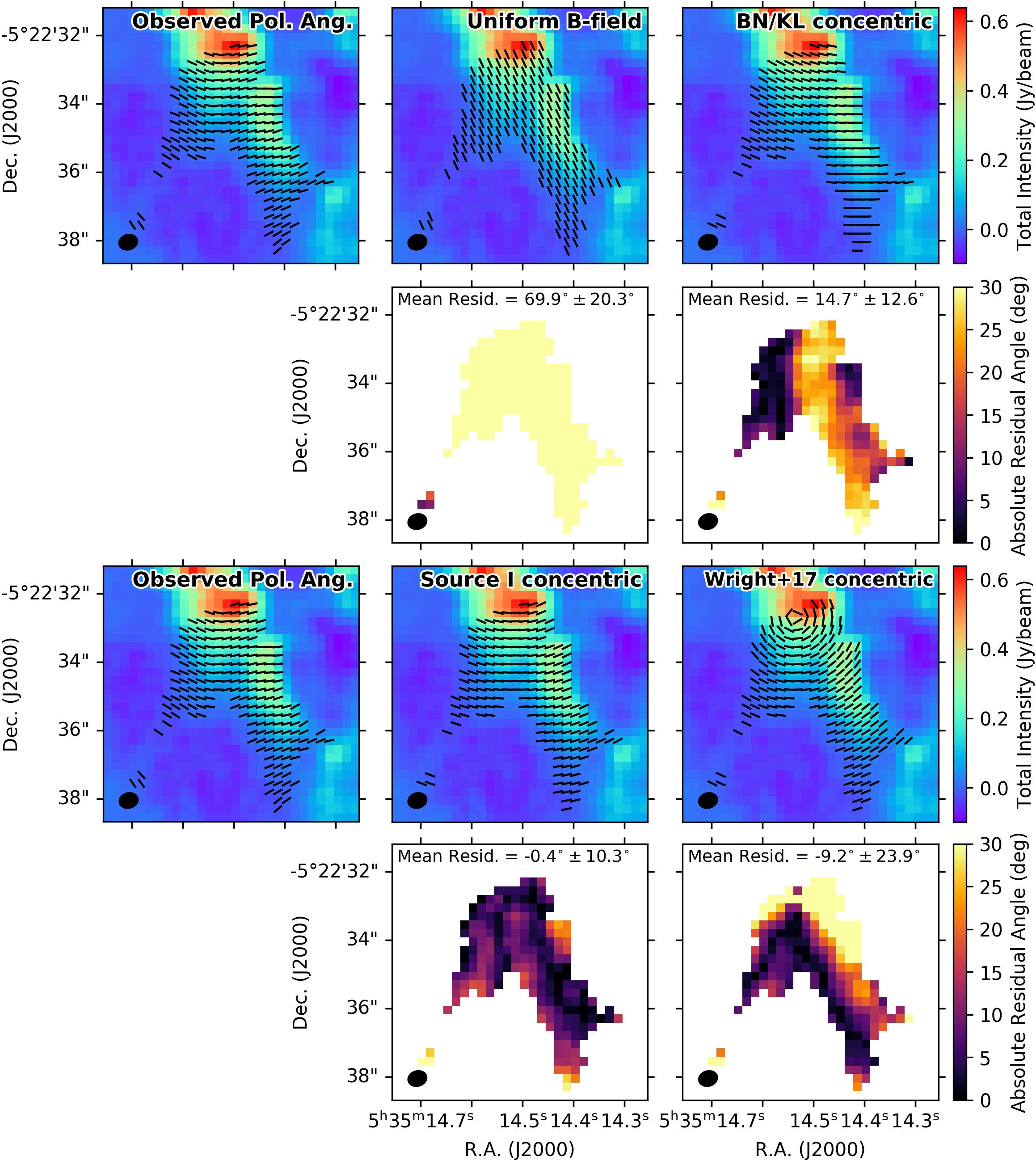}
    \caption{Comparison of models in the Anomalous Region/Fork. 
    {First and third rows show} observed and model polarisation geometries, plotted on Stokes $I$ emission, {second and fourth} show absolute difference in angle between data and models.  {Left column shows observed polarisation vectors.  Central column shows polarisation vectors perpendicular to the large-scale magnetic field direction (top; alignment mechanism: B-RATs) and concentric around Source I (bottom; hypothesised alignment mechanism: k-RATs).  Right column shows polarisation vectors concentric around the BN/KL explosion centre (top; hypothesised alignment mechanism: B-RATs/v-MATs) and concentric around the centre of the \citet{wright2017} ring feature (bottom; hypothesised alignment mechanism: v-MATs).  In the second and fourth rows, the colour table saturates at a difference in angle of $30^{\circ}$, to emphasise the differences between the models.}  All maps are shown on 0.25\arcsec\ (approximately Nyquist-sampled) pixels.  The synthesised beam size is shown in the lower left-hand corner of each plot.}
    \label{fig:compare_AR}
\end{figure*}

There is a significant `fork' visible to the south of the Hot Core in the polarised intensity and polarisation angle maps, as shown in Figures~\ref{fig:PI} and \ref{fig:pa}.  Vectors on the western side of the fork have a typical polarisation angle significantly different both to that across the Ridge and to that in the eastern arm.
\citet{rao1998}, observing OMC-1 in polarised light with BIMA, identified this as an `anomalous region', with polarisation vectors significantly different from elsewhere in OMC-1, and inconsistent with {being perpendicular to} the large-scale field direction.  The vectors which we see are consistent with their observations.  \citet{rao1998} suggested that grains in this region are mechanically aligned by the Gold effect, driven by the Source I outflow\footnote{{The `outflow' referred to by \citet{rao1998} is now recognized to be high-velocity gas from the BN/KL explosive event.}}.

{\citet{wright2017} suggest that much of the dust emission in the Fork originates from the walls of the cavity formed by the bipolar outflow from Source I, based on the spatial coincidence of dust emission and SiO emission tracing the outflow.  The Source I outflow is shown in Figure~\ref{fig:PI}  In the following discussion we consider both the case in which emission arises from the outflow cavity wall, and that in which it arises from the ambient medium of OMC-1.}

\subsubsection{Comparison of data and models}
\label{sec:fork_ks}

\begin{figure}
    \centering
    \includegraphics[width=0.47\textwidth]{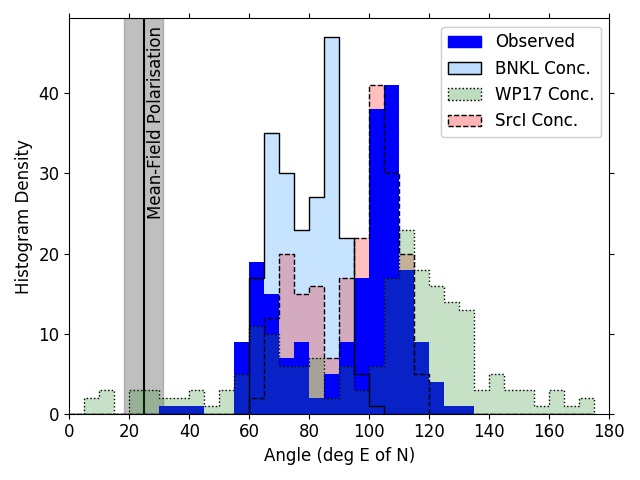}
    \caption{Histogram of observed polarisation angles in the Anomalous Region/Fork (blue), compared with models: polarisation vectors concentric around (1) the centre of the BN/KL explosion (light blue, solid outline), (2) the centre of the \citet{wright2017} ring feature (green, dotted outline), (3) Source I (red, dashed outline).  The polarisation angle associated with the mean 116-degree magnetic field direction is marked.  Polarisation angles are {measured on 0.25\arcsec\ (approximately Nyquist-sampled) pixels and} shown in the range 0--180 degrees for clarity.}
    \label{fig:histogram_AR}
\end{figure}

Figures~\ref{fig:compare_AR} and \ref{fig:histogram_AR} show that the polarisation vectors in the Fork are inconsistent with the $26^{\circ}$ polarisation direction associated with the large-scale magnetic field, but broadly consistent with being concentric around either the BN/KL explosion centre, the centre of a `ring feature' identified by \citet{wright2017} (discussed below), or Source I.  Figure~\ref{fig:compare_AR} shows each of these polarisation geometries, along with the absolute residual angles between the models and the observed polarisation geometry.  The BN/KL-concentric model is {broadly} consistent with the observations in the eastern arm of the Fork, but systematically different by $\sim 20^{\circ}$ in the western arm.  The ring-feature-concentric model is consistent with the observations in the south of the region, but not consistent in the north; we discuss this further below. The Source I-concentric model is broadly consistent with the observed polarisation geometry {in the western arm of the Fork}. 

{We quantified the similarity of these models to the data using two-sided Kolmorogov-Smirnov (KS) and Kuiper tests.  We chose the KS test as the most widely-used statistic for comparing the similarity of two distributions, and the Kuiper test as a more appropriate measure of similarity for distributions of cyclic variables \citep[e.g.][]{aizawa2020}.}

\begin{figure*}
    \centering
    \includegraphics[width=\textwidth]{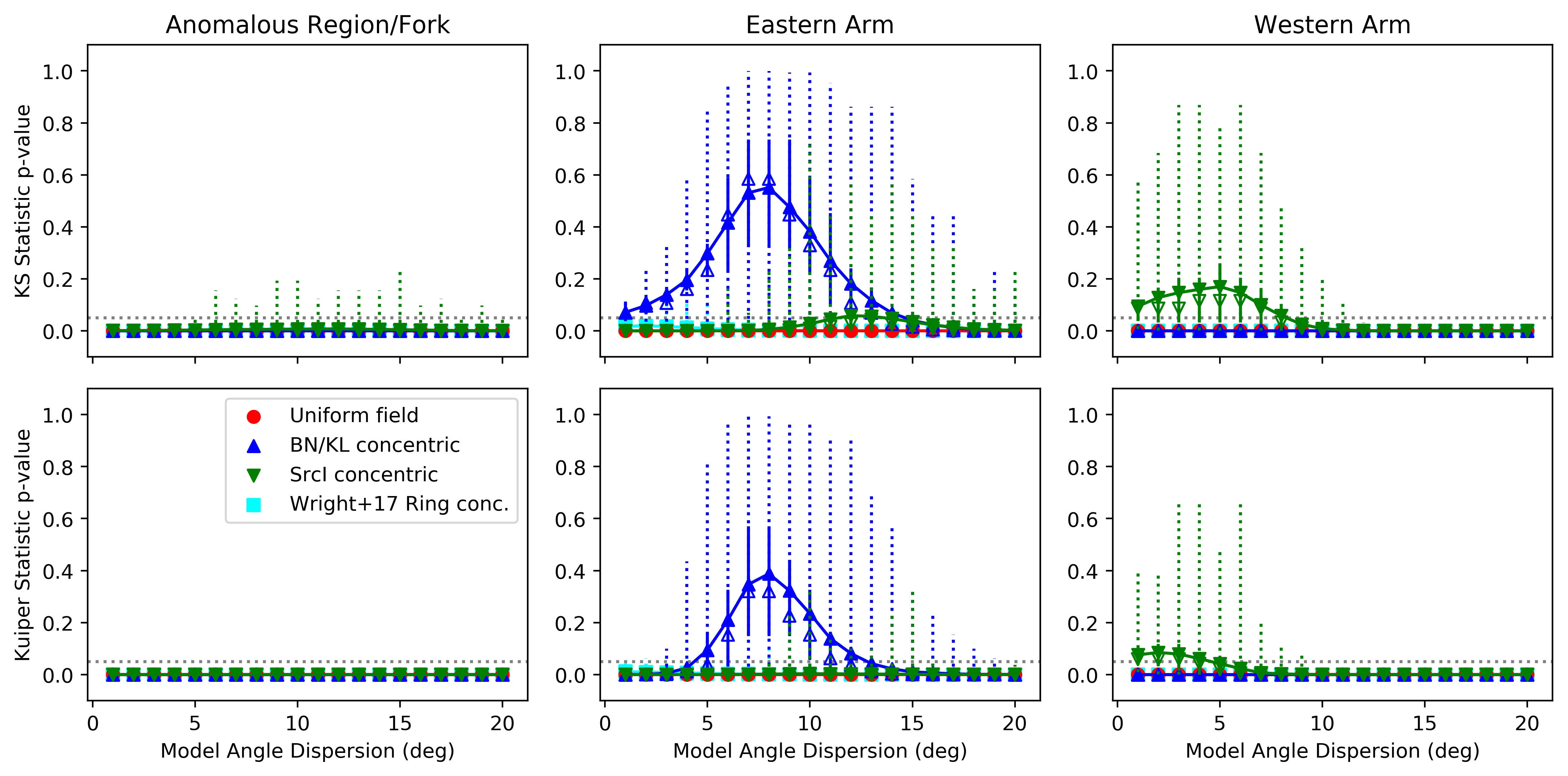}
    \caption{{Results of Kolmogorov-Smirnov (KS) and Kuiper tests comparing observed polarisation angles in the Anomalous Region/Fork to a set of realisations of our four models with specified values of Gaussian dispersion in polarisation angle.  Top row shows the $p$ values associated with the KS statistic and bottom row shows those associated with the Kuiper statistic.  Left column shows results for the entire Fork; centre column, for the eastern arm; and right column, for the western arm.  In each panel, solid symbols show the mean value, open symbols, the median.  Dotted error bars show the full range of the $p$-value across the set of realisations; solid error bars show the interquartile range.  The dashed grey line marks the $p=0.05$ criterion for statistical significance.}}
    \label{fig:kstest_AR}
\end{figure*}

{For each model, we determined polarisation angles predicted in each pixel by the model, and drew sets of perturbations on these angles from a Gaussian distribution of width $\sigma_{\theta}$.  We then performed two-sided KS and Kuiper tests comparing the model, with added dispersion in angle, to the data.  We repeated this process 1000 times for each value of $\sigma_{\theta}$ considered.  We tested $\sigma_{\theta}$ values in the range $1^{\circ} - 20^{\circ}$.  Additional dispersion in angle over our measured uncertainties ($\delta\theta < 3^{\circ}$) could result from Alfv\'enic distortion of the magnetic field by non-thermal gas motions \citep{davis1951a,chandrasekhar1953}.  The results of these tests are shown in Figure~\ref{fig:kstest_AR}.} 

{According to both tests, the BN/KL-concentric model is the only model that can be made consistent ($p > 0.05$) with the data in the eastern arm, while the Source-I-concentric model is the only model that can be made consistent with the data in the western arm.  The two tests produce similar results, with the Kuiper test generally returning a more narrow range of angles over which the model and the data agree with a probability $p > 0.05$.}

{In the eastern arm, the observed polarisation geometry is consistent with the \citet{cortes2020} model in which the magnetic field is radial around BN/KL; however, additional scatter in the observed polarisation angle above our observed uncertainty must be introduced.  The best agreement between the BN/KL-concentric model and our data occurs for $\sigma_{\theta} = 8^{\circ}$, but $p > 0.05$ agreement occurs up to $\sigma_{\theta} \approx 13^{\circ}$ (KS test), or in the range $4.5^{\circ}\lesssim \sigma_{\theta}\lesssim 12.5^{\circ}$ (Kuiper test).}

{In the western arm, $p > 0.05$ agreement between the Source-I-concentric model and our data occurs at $\sigma_{\theta} < 8^{\circ}$ (KS test), or $\sigma_{\theta} < 5^{\circ}$ (Kuiper test).  The difference in modelled angle dispersion between the eastern and western arms of the Fork -- and between the western arm and elsewhere in OMC-1, as described below -- is suggestive either of different non-thermal gas motions within the western arm, or of a different alignment mechanism between the two arms.}

We consider {five} hypotheses to explain the observed polarisation pattern {in the western arm}: (1) grains aligned by B-RATs with respect to a distorted magnetic field; {(2)} polarisation arising from scattering of emission from Source I; {(3)} polarisation arising from supersonic mechanical alignment (Gold alignment) induced by either the BN/KL explosion or the Source I outflow; {(4)} grains aligned by subsonic v-MATs, induced by (a) the shock associated with the passage of ejecta from the BN/KL explosion through the region; (b) shocks associated with the Source I outflow; {(5)} grains aligned by {k-RATs}, perpendicular to the radiation gradient associated with Source I.  We consider these hypotheses in turn below, {and then as a check on our analysis confirm that the gas damping timescale is sufficiently long in the region to allow a preferential dust precession axis to exist.  The following discussion is summarised in Table~\ref{tab:fork}.}

\begin{table*}
    \centering
    \begin{tabular}{l l l l}
    \hline
            & Characteristic & Timescale Value & Dominant Grain Size Regime\\
    Mechanism & Timescale & (years) & ($\mu$m) \\
    \hline
    B-RATs & $\tau_{Lar}$ & $\sim 1.6(10^{-5} -10^{-3})\,a_{-5}^{2}$ & $\lesssim 0.005 - 0.1$ \\
    k-RATs & $\tau_{rad,p}$ & $\lesssim 1.7\times 10^{-5} \,a_{-5}^{\frac{1}{2}}$ & $\gtrsim 0.005 - 0.1$ \\
    v-MATs & $\tau_{mech}$ & $\gtrsim 72 (\omega/\omega_{th})$ & --- \\
    Randomisation & $\tau_{gas}$ & $\sim 0.017 - 1.7\,a_{-5}$ & --- \\
    \hline
    \end{tabular}
    \caption{A summary of the timescales which we estimate in the Anomalous Region/Fork.  Note that $a_{-5} = (a/10^{-5}\,{\rm cm})$; $a_{-5} = 1$ indicates grain size $a = 0.1\,\mu$m.}
    \label{tab:timescales}
\end{table*}

\subsubsection{Distorted magnetic field}

As shown in Figures~\ref{fig:compare_AR} and \ref{fig:histogram_AR}, the polarisation pattern in the {western arm of the} Fork is inconsistent with that predicted based on the large-scale mean field direction {and with the radial field geometry observed elsewhere in OMC-1 by \citet{tang2010} and \citet{cortes2020}}.  \citet{tang2010} hypothesised that the polarisation geometry of OMC-1, observed at 870$\mu$m using the SMA, arose from grains aligned perpendicular to a magnetic field which had been significantly distorted from its initial configuration.  They proposed two hypotheses for how the field had been distorted: (1) that the observed polarisation indicated that the density structure of the centre of OMC-1 forms a rotating `pseudo-disk' around the centre of the BN/KL explosion, with a toroidal magnetic field.  We consider this hypothesis to have been disfavoured by ALMA studies better determining the line-of-sight distances and velocities of the various OMC-1 clumps \citep{pagani2017}.  (2) that the BN/KL explosive outflows have realigned the magnetic field to be radial around the explosion centre (i.e. polarisation vectors are concentric around the BN/KL centre).  {Our observations suggest that the vectors in OMC-1, including in the eastern arm of the Fork, are mostly consistent with being concentric around the BN/KL explosion centre.}  However, {in the western arm, the polarisation geometry which we observe is more consistent with being concentric around Source I, suggesting that if the grains remain aligned with the magnetic field in the region, the field is likely to be radial around that source.}  This could potentially indicate a highly poloidal field in the Source I outflow, {or the magnetic field being well-coupled to infalling or outflowing gas}.

\subsubsection{Larmor timescale}
\label{sec:fork_tlar}

We will consider alternative explanations for the observed polarisation geometry in the {western arm of the} Fork, which do not require such wholesale reorganisation of the magnetic field.  For dust grains to be aligned with respect to the magnetic field direction, rather than some other axis, the timescale for precession around the magnetic field direction (the Larmor timescale, $\tau_{Lar}$) must be shorter than all other precession timescales \citep[e.g.][]{hoang2016}.  Therefore we estimate $\tau_{Lar}$ in the {western arm of the} Fork, for comparison with other timescales.

The Larmor timescale is given by \citet{tazaki2017} as
\begin{equation}
    \tau_{Lar} \simeq 1.3\hat{\rho}\hat{s}^{-\frac{2}{3}}a_{-5}^{2}\hat{B}^{-1}\hat{\chi}^{-1}\,\,{\rm year},
    \label{eq:tlar}
\end{equation}
where $\hat{\rho} = \rho/3$\,g\,cm$^{-3}$ and $\rho$ is the mass density of the grains, $\hat{s} = s/0.5$ and $s$ is the axial ratio of the dust grains, $a_{-5} = a/10^{-5}$\,cm and $a$ is the radius of the dust grains, $\hat{\chi} = \chi(0)/10^{-4}$ and $\chi(0)$ is the zero-frequency magnetic susceptibility of the grains, and $\hat{B} = B/5\mu$G, and $B$ is magnetic field strength.  We take $\hat{\rho}\sim 1$ and $\hat{s}\sim 1$.

{Measurements made on larger scales in the region indicate a magnetic field strength $B\lesssim 10$\,mG at densities $n_{\textsc{h}}\sim 10^{6}$\,cm$^{-3}$ \citep{hildebrand2009,houde2009,pattle2017}, similar to the ambient density in which Source I is moving \citep{wright2017}.  \citet{cortes2020} find $B = 9.4 \pm 1.8$ mG at $n_{\textsc{h}_{2}} = 2.7\times 10^{8}$\,cm$^{-3}$. The similarity in gas densities between the Fork and the region of OMC-1 in which this magnetic field strength was determined \citep{favre2011} and the strong coupling between the magnetic field and the gas in OMC-1 \citep{cortes2020} suggest that this value is likely to be representative of the field strength in the Fork, and so we adopt $\hat{B}\sim 2000$.  All of these estimates of magnetic field strength were made using variants of the Davis-Chandrasekhar-Fermi (DCF) method \citep{davis1951a,chandrasekhar1953}.}

\citet{draine1996} gives values for magnetic susceptibility $\chi(0)$ in the range $4.2\times 10^{-5}-4.2\times10^{-3}$ for paramagnetic grains, and so we take $\hat{\chi}\sim 0.4-40$.  These combine to give
\begin{equation}
    \tau_{Lar}\sim (1.6\times 10^{-5} - 1.6\times 10^{-3})\,a_{-5}^{2}\,\,{\rm year}.
    \label{eq:tlar_emp}
\end{equation}
{\citet{cortes2020} note that their derived magnetic field strength, $B = 9.4 \pm 1.8$ mG, is effectively an upper limit on the true magnetic field strength, due to the beam- and line-of-sight-averaging effects inherent in the DCF method (see, e.g. \citealt{pattle2019} for a discussion).  Thus, the value of $\tau_{Lar}$ given in equation~\ref{eq:tlar_emp} is a lower limit, for our adopted range of $\chi$ values.  However, we note that if the dust grains in the Fork were to have super-paramagnetic inclusions \citep[e.g.][and refs. therein]{lazarian2019}, $\chi$ could be made considerably larger, correspondingly decreasing $\tau_{Lar}$.  The uncertainty on $\tau_{Lar}$ is thus difficult to quantify: \citet{cortes2020} give a formal uncertainty on $B$ of $\sim 20$\%, but this is dwarfed by the two orders of magnitude (or larger) uncertainty on $\chi$ \citep{draine1996}, and so we consider only the latter when estimating the range of plausible values of $\tau_{Lar}$.}

\subsubsection{Dust self-scattering}

The polarisation pattern in the {western arm of the} Fork is consistent with being concentric around Source I.  This could imply that polarisation arises from scattering of light from Source I \citep{kataoka2015}.  We note however that as is the case in the Source I disc, this would require grain sizes $\sim 140\mu$m.  If the emission from the {western arm of the} Fork arises from the ambient medium of OMC-1, this level of grain growth appears impossible.  If the emission arises from the Source I outflow cavity, and if such large grains exist in the Source I disc as is suggested by the Source I polarisation pattern, it could be hypothesised that they might be entrained into the outflow.  However, transport of such large dust grains, as well as their avoiding destruction in the outflow in sufficient number to produce the observed polarisation pattern, does not seem likely \citep{giacalone2019}.  We thus discount this hypothesis, while noting that we cannot definitively rule it out.

\subsubsection{Supersonic mechanical (Gold) alignment}

For completeness, we note the possibility of supersonic mechanical alignment (Gold alignment; \citealt{gold1952}).  If the polarised emission in the Fork arises from the ambient medium of OMC-1, and is associated with BN/KL shocks, Gold alignment would produce a radial polarisation pattern around the centre of the BN/KL explosion.  If the emission in the {western arm of the} Fork arises from the Source I outflow cavity walls, Gold alignment would produce polarisation parallel to the Source I outflow, or radial around Source I.  All of these geometries are inconsistent with the observed polarisation pattern shown in Figure~\ref{fig:pa}, and so we do not consider Gold alignment further.

\subsubsection{Mechanical Alignment Torques}

\citet{lazarian2007b} proposed the Mechanical Alignment Torques (MATs) mechanism, in which grains drifting relative to gas are aligned by mechanical torques to have their long axes perpendicular to the precession axis of the grain.  This precession axis is typically the magnetic field direction \citep[B-MATs,][]{lazarian2007b,hoang2018}, but can in some environments be the velocity vector of the gas/dust drift \citep[v-MATs,][]{lazarian2007b,hoang2016}.
The v-MAT alignment mechanism can occur when the velocity difference between the gas and dust is subsonic, and when the mechanical alignment timescale (the precession time around the gas flow), $\tau_{mech}$, is shorter than the Larmor precession timescale, $\tau_{Lar}$ (cf. \citealt{hoang2018}, Sec. 6.4).  This mechanism further requires the dust grains to have significant helicity, which is acquired through coagulation.
\citep{brauer2008,ormel2009,hirashita2012}.

\citet{hoang2018} discuss environments where gas/dust drift is likely to be induced, concluding that such drift may be triggered by cloud-cloud collisions, radiation pressure, ambipolar diffusion, or gravitational sedimentation.  Gas/dust drift occurs across shock fronts \citep{mckee1987}, and so can plausibly expected to be occurring in OMC-1, either in the aftermath of the BN/KL explosion, or in shocks within the Source I outflow.  Shocks in OMC-1 are thought to be continuous (C-shocks; e.g. \citealt{colgan2007}), supporting the hypothesis that the magnetic field in the region is dynamically important \citep{draine1980}.

\begin{figure*}
    \includegraphics[width=\textwidth]{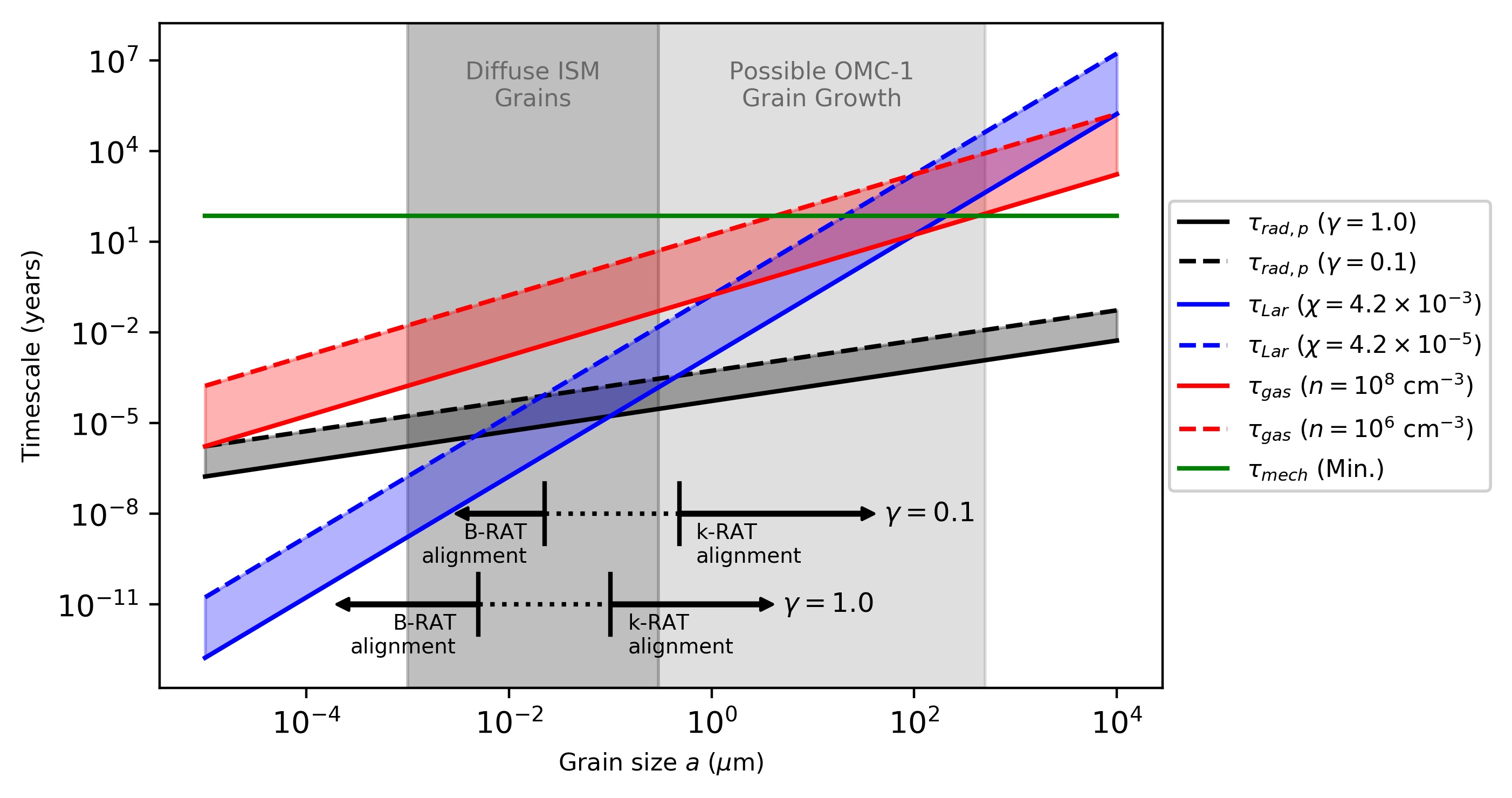}
    \caption{A comparison of the radiative precession ($\tau_{rad,p}$), Larmor ($\tau_{Lar}$), gas damping ($\tau_{gas}$) and mechanical alignment ($\tau_{mech}$) timescales which we estimate in the {western arm of the Anomalous Region/Fork}, as a function of grain size $a$.  Note that the shortest timescale determines the precession axis; thus, small grains are {likely to be} aligned by B-RATs to precess around the magnetic field direction, and large grains {are likely to be aligned} by k-RATs to precess around the radiation anisotropy vector.  The minimum value of $\tau_{mech}$ which we show assumes $\omega\approx\omega_{th}$, i.e. that grains are rotating at the thermal angular velocity.  {Other timescales are shown at their fastest and slowest likely values as determined by the dominant quantifiable source of uncertainty in their determination, as discussed in the text, with their likely range shaded.}  The dark grey shaded region marks the grain size distribution in the diffuse ISM \citep{draine2007}; the light grey shaded area shows the maximum extent of grain growth likely in OMC-1 ($a_{max}<500\,\mu$m; see Section~\ref{sec:srcI}), although we note that grains larger than a few microns are unlikely to be found outside of the Source I disc.}
    \label{fig:timescales}
\end{figure*}

If the emission in the Fork arises from the ambient medium of OMC-1, rather than from the Source I outflow cavity walls, we might expect grains to be mechanically aligned by shocks associated with the BN/KL explosion ejecta.
While the observed polarisation geometry is, in the western arm of the Fork, inconsistent with being concentric around the BN/KL explosion centre, \citet{wright2017} identified a ring of emission near SMA 1 in HCN 354.5\,GHz and H$_{3}$CN  354.7\,GHz emission, which they interpreted as evidence for passage of debris from the BN/KL explosion.  The ring has $\sim$2\,km\,s$^{-1}$ expansion velocity and a dynamical age of $\sim$700 yr, consistent with the approximate age of the BN/KL explosion, if its expansion has been somewhat decelerated.  {This feature is at a different systemic velocity ($+12$\,km\,s$^{-1}$) than is the material which \citet{wright2017} associate with the Source I outflow (located at $-12$ to $-7$\,km\,s$^{-1}$), and is likely to be located behind the BN/KL explosion centre.}  As well as considering grain alignment concentric around Source I, we consider grain alignment concentric around the centre of the ring, in case mechanical alignment were induced by the shock associated with these particular ejecta.  In the south of the Anomalous Region/Fork (i.e. at larger radii), the polarisation pattern is more consistent with being concentric around the position of the ring than it is with being concentric around the BN/KL explosion centre, but the model fails at positions near the ring centre.

Alternatively, if the polarised emission in the Fork arises from the Source I outflow cavity walls, grains cannot be aligned by shocks associated with the BN/KL explosion, as Source I and its associated outflow is moving behind these shock fronts \citep[e.g.][]{hirota2020}.  In this case, v-MAT alignment could instead be induced by shocks associated with the expansion of the bipolar outflow into its surroundings.  Polarisation vectors might then be expected to be radial around Source I, perpendicular to the surface of the outflow, or less ordered, depending on the nature of the outflow shocks.

\begin{figure*}
    \centering
    \includegraphics[width=\textwidth]{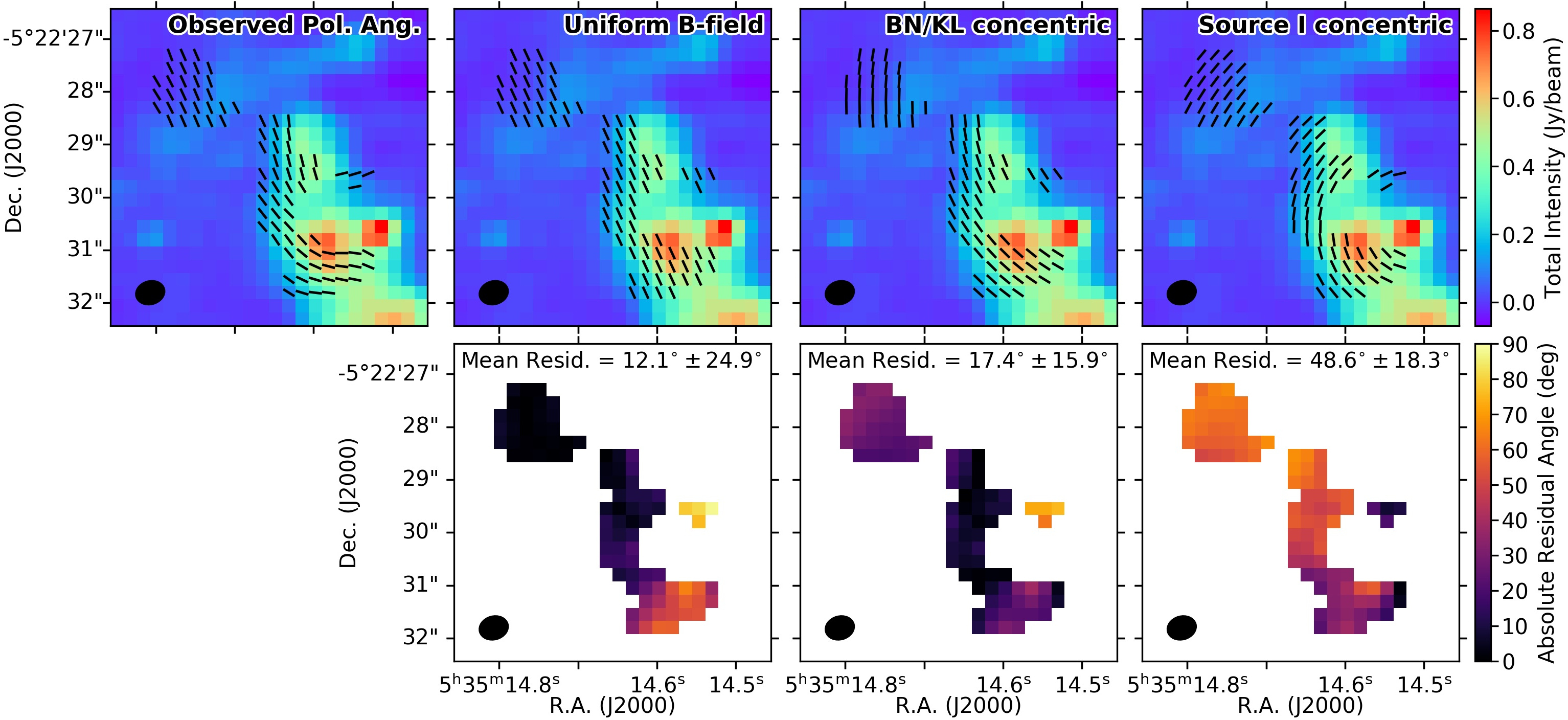}
    \caption{Comparison of models in the Ridge.  Top row shows observed and model polarisation geometries, plotted on Stokes $I$ emission, bottom row shows absolute difference in angle between data and models.  Far left: Observed polarisation vectors.  Centre left: polarisation vectors aligned $26^{\circ}$ E of N, perpendicular to the large-scale magnetic field direction (hypothesised alignment mechanism: B-RATs).  Centre right: polarisation vectors concentric around the BN/KL explosion centre (hypothesised alignment mechanism: {B-RATs/v-MATs}).  {Far right}: polarisation vectors concentric around Source I (hypothesised alignment mechanism: k-RATs).  All maps are shown on 0.25\arcsec\ (approximately Nyquist-sampled) pixels.  The synthesised beam size is shown in the lower left-hand corner of each plot.}
    \label{fig:compare_Ridge}
\end{figure*}

\subsubsection{Mechanical alignment timescale}

The timescale for alignment by v-MATs is given by
\begin{equation}
    \tau_{mech} \simeq 36\left(\frac{c_{s}}{\Delta v}\right)^{2}\left(\frac{\omega}{\omega_{th}}\right)\hat{s}^{2}\frac{1}{\sin 2\Theta}\,{\rm year},
    \label{eq:tmech}
\end{equation}
(Lazarian \& Hoang, ApJ subm.), where $c_{s}$ is gas sound speed, $\Delta v$ is gas/dust velocity difference, $\omega$ is the grain angular velocity, $\omega_{th}$ is the thermal angular velocity, and $\Theta$ is the angle between the grain axis of major inertia and the direction of radiation.
We take $\hat{s}\sim1$ and $\sin 2\Theta \sim 0.5$.

The velocity difference between gas and dust in C-type shocks is not well-characterised, potentially taking any value between zero and the shock velocity, depending on environment  \citep{wardle1998, guillet2007}.  However, the condition for v-MATs is $\Delta v < c_{s}$, and so equation~(\ref{eq:tmech}) becomes
\begin{equation}
    \tau_{mech}\gtrsim 72 \left(\frac{\omega}{\omega_{th}}\right)\,{\rm year}.
    \label{eq:tmech_emp}
\end{equation}

Comparison of equations~\ref{eq:tlar_emp} and \ref{eq:tmech_emp} suggests that $\tau_{mech} \gg \tau_{Lar}$.  The requirement for $\tau_{mech}<\tau_{Lar}$ to hold is the physically implausible condition $\omega/\omega_{th}\ll 1$, i.e. grains would have to be rotating subthermally.  The timescale for v-MAT alignment thus remains too long for this mechanism to be likely to be the main cause of grain alignment in the Fork.

\subsubsection{Radiative Alignment Torques}

Under the Radiative Alignment Torques (RATs) paradigm of grain alignment, grains are efficiently aligned
when they can be spun up to suprathermal rotation by
an anisotropic radiation field \citep{dolginov1976,lazarian2007a}.  As with MATs, the grains will align with their long axes perpendicular to their precession axis.  In the large majority of ISM environments, the precession axis can be presumed to be the magnetic field direction \citep[B-RATs][]{lazarian2007a}.  However, in the presence of a sufficiently strong and anisotropic radiation field, the precession axis can instead be the radiation anisotropy vector, and so grains will be aligned with their major axes concentric around the source driving the radiation field \citep[k-RATs; ][]{tazaki2017}.  The condition for k-RATs to dominate over B-RATS is that the radiative precession timescale must be shorter than the Larmor timescale, i.e. $\tau_{rad,p} < \tau_{Lar}$ \citep{lazarian2007a,tazaki2017}.  {Alignment by k-RATs has not been definitively observed outside of protostellar discs; however, there has been a recent tentative detection in HAWC+ observations of the nearby Orion Bar \citep{chuss2019}, and a potential detection in the envelope of an evolved star \citep{andersson2018}}.

The brightest source in OMC-1 is Source I, with a luminosity $> 10^{4}$\,L$_{\odot}$ \citep{menten1995}.  As shown in Figure~\ref{fig:compare_AR}, the polarisation pattern in the Fork is quite consistent with being concentric around Source I, potentially suggesting that the dust grains in the region are aligned by k-RATs, driven by the radiation field of Source I.  In the following section we estimate the radiative precession timescale $\tau_{rad,p}$ {arising from the unobscured radiation field of Source I} in the Anomalous Region/Fork.
We do not include other sources of radiation in OMC-1 in this analysis, as it is the strongly anisotropic radiation field of Source I which we hypothesise is driving k-RAT alignnment in the Fork.

\subsubsection{Radiative precession timescale}

The radiative precession timescale is given by \citet{tazaki2017} as
\begin{equation}
    \tau_{rad, p}\simeq 110\,\hat{\rho}^{\frac{1}{2}}\hat{s}^{-\frac{1}{3}}a_{-5}^{\frac{1}{2}}\hat{T}_{d}^{\frac{1}{2}}\left(\frac{u_{rad}}{u_{\textsc{isrf}}}\right)^{-1}\left(\frac{\bar{\lambda}}{1.2\,\mu{\rm m}}\right)^{-1}\left(\frac{\gamma|Q_{\Gamma}|}{0.01}\right)^{-1}\,\,{\rm year},
    \label{eq:trad}
\end{equation}
where $\hat{T}_{d} = T_{d}/15\,{\rm K}$ and $T_{d}$ is dust temperature; $u_{rad}$ is the energy density of the radiation field in the region under consideration; $u_{\textsc{isrf}}$ is the energy density of the standard interstellar radiation field (ISRF), given by \citet{tazaki2017} as $u_{\textsc{isrf}} = 8.64\times 10^{-13}$\,erg\,cm$^{-3}$; $\bar{\lambda}$ is the {mean wavelength of the incident radiation spectrum}, $\gamma$ is radiation field anisotropy, and $|{Q_{\Gamma}}|$ is the RAT efficiency.  This formulation of $\tau_{rad, p}$ assumes grains to be rotating at the thermal angular velocity, i.e. $\omega\approx\omega_{th}$ \citep{lazarian2007a}.

We again take $\hat{\rho}\sim 1$ and $\hat{s}\sim 1$.  We expect $0.1< \gamma \leq 1$, as $\gamma \sim 0.1$ in the diffuse ISM \citep{draine1996a}, and $\gamma \lesssim 1$ in the immediate vicinity of a protostar \citep{tazaki2017}.  As we are specifically considering the radiation field from Source I, which we expect to be strongly anisotropic, we take $\gamma\sim 1$. {Note that this implies that radiation from Source I is effectively unobscured in the western arm of the Fork; a justifiable assumption if the polarised emission indeed arises from the Source I outflow cavity wall, which we discuss further below.}  We further take $|{Q_{\Gamma}}| \leq 0.4$ \citep{lazarian2007a,tazaki2017}, and so $\gamma|{Q_{\Gamma}}|/0.01 \lesssim 40$.

The luminosity of Source I is not well-characterised, but is thought to be $> 10^{4}$\,L$_{\odot}$ \citep{menten1995}.  The plane-of-sky separation between the Fork and Source I is $\sim 3.5$\arcsec, which at a distance of 388\,pc corresponds to {$\sim 2\times 10^{16}$\,cm} ($\sim 1400$\,au).  We thus estimate the radiation energy density in the vicinity of the Fork to be
\begin{equation}
    u_{rad} = \frac{L}{4\pi R^{2}c} > 2.7\times 10^{-7}\,{\rm erg\,cm}^{-3},
    \label{eq:urad}
\end{equation}
where $L$ is the luminosity of Source I and $R$ is the separation between Source I and the Fork.

The effective brightness temperature of Source I is $\sim 1500$\,K \citep{reid2007}, and so from Wien's Law, we infer a peak emission wavelength {of photons emanating from Source I} of $\sim 1.9\,\mu$m.  We thus take $\bar{\lambda}/1.2\mu{\rm m}\sim 1.6$.  {We note that as with taking $\gamma \sim 1$, this assumes that emission from Source I is unobscured in the Fork.  We discuss this assumption further below.}

Dust temperature $T_{d}$ can be estimated for silicates using the relation
\begin{equation}
    T_{d}\approx 16.4\left(\frac{u_{rad}}{u_{\textsc{isrf}}}\right)^\frac{1}{6}\,{\rm K}
    \label{eq:Td}
\end{equation}
{\citep{draine2011}.  Using our value of $u_{rad}$ from equation~(\ref{eq:urad}), we estimate $T_{d}\approx 135$\,K in the Fork, and so $\hat{T}_{d}^{\frac{1}{2}}\approx 3$}.

Combining these estimates, equation~(\ref{eq:trad}) becomes
\begin{equation}
    \tau_{rad, p} \lesssim 1.7\times 10^{-5} \,a_{-5}^{\frac{1}{2}}\,\,{\rm year}.
    \label{eq:trad_emp}
\end{equation}

Comparing this to the Larmor timescale in the Fork, as given in equation~(\ref{eq:tlar_emp}), we find the condition for k-RATs to dominate over B-RATs, $\tau_{rad,p} < \tau_{Lar}$, is equivalent to
\begin{equation}
    a_{-5} > 0.05 - 1.0,
\end{equation}
or equivalently,
\begin{equation}
    a > 0.005 - 0.1\,\mu{\rm m}.
\end{equation}
This suggests that in the vicinity of Source I, $\tau_{rad,p} < \tau_{Lar}$ will hold for larger paramagnetic dust grains {if Source I remains relatively unobscured}, and so we can plausibly expect to see a polarisation pattern arising from k-RATs.

While highly uncertain, the minimum values of $a$ for which $\tau_{rad,p} < \tau_{Lar}$ that we find are plausible grain sizes in a dense molecular cloud \citep[e.g.][]{draine2007}.  The maximum grain size in the diffuse ISM is $0.25-0.3\,\mu$m \citep{mathis1977,draine2007}, indicating that while k-RATs could potentially dominate over B-RATs in the vicinity of Source I even in relatively pristine ISM material, the grain growth which is likely to have occurred in such a dense environment \citep{ysard2013} makes $\tau_{rad,p} < \tau_{Lar}$ more likely to hold.  Moreover, if grains are indeed aligned by k-RATs downstream of the shocks associated with BN/KL ejecta and/or the expansion of the Source I outflow, it suggests that these shocks have not destroyed all of the larger dust grains in the cloud.

\subsubsection{Gas damping timescale}
\label{sec:fork_tgas}

A further requirement for grains to precess around any given axis is that the precession timescale around that axis is shorter than the gas damping timescale $\tau_{gas}$, the characteristic timescale of grain randomisation by gas collisions \citep{lazarian2007a}.  The highly ordered polarisation geometry of the Fork -- and across OMC-1 -- strongly suggests that the dust grains are not randomised.  However, as a check on our previous analysis, we estimate the gas damping timescale in the Fork.

{\citet{hoang2016} give $\tau_{gas}$ as}
\begin{equation}
    \tau_{gas} = 6.6\times10^{4}\, \hat{\rho}\hat{s}^{-\frac{2}{3}}a_{-5}\Gamma_{\parallel}^{-1} \left(\frac{300\, {\rm K}^{\frac{1}{2}}{\rm cm}^{-3}}{T_{gas}^{\frac{1}{2}}n_{\textsc{h}}}\right)\,\,{\rm year},
    \label{eq:tgas}
\end{equation}
where $\Gamma_{\parallel}$ is a factor of order unity characterising grain geometry and $T_{gas}$ is gas temperature.  We take $\Gamma_{\parallel}\sim 1$, continue to take $\hat{s}\sim 1$ and $\hat{\rho}\sim 1$ and {$n_{\textsc{H}}\sim 10^{6} - 10^{8}$\,cm$^{-3}$ (cf. Section~\ref{sec:fork_tlar})}, and assume $T_{gas} \sim T_{dust} \approx 135$\,K.  Equation (\ref{eq:tgas}) thus becomes
\begin{equation}
    \tau_{gas}\sim 0.017 - 1.7\,a_{-5}\,\,{\rm year}.
    \label{eq:tgas_emp}
\end{equation}
Comparison of equation~(\ref{eq:tgas_emp}) with equations~(\ref{eq:tlar_emp}) and (\ref{eq:trad_emp}) shows that there is no value of $a_{-5}$ at which $\tau_{gas}$ is the shortest timescale.  $\tau_{gas}< \tau_{rad,p}$ holds only for unphysically small grains, with $a_{-5} < 10^{-10}$, at which size $\tau_{Lar} \ll \tau_{gas}$ would hold if such grains existed.  Conversely, $\tau_{gas}< \tau_{Lar}$ only for unphysically large grains, with $a_{-5} > 10^{3}-10^{5}$, at which size $\tau_{rad,p} \ll \tau_{gas}$.  These timescales are summarised in Table~\ref{tab:timescales} and illustrated in Figure~\ref{fig:timescales}.   For the values of $\tau_{Lar}$ and $\tau_{rad,p}$ which we find in the Fork, $\tau_{gas}$ would need to be smaller by at least {three to} five orders of magnitude for a regime to exist in which it is the shortest timescale.  Thus grains in the Fork cannot have their alignments randomised by gas collisions faster than they can be induced to precess around either the magnetic field direction or the radiation anisotropy gradient by RATs.

\subsubsection{Discussion of grain alignment in the Anomalous Region/Fork}
\label{sec:fork_discuss}

In the preceding analysis, {by taking $\gamma\sim 1$ and $\bar{\lambda} = 1.9\,\mu$m} we have assumed that the polarised emission arises in a location where there is minimal obscuration of Source I.  Such obscuration would introduce absorption, re-emission and scattering of radiation, reducing the anisotropy $\gamma$ in the radiation field and increasing $\bar{\lambda}$, and so increasing $\tau_{rad,p}$.  {If $\gamma < 1$ but remains above the value in the diffuse ISM, the minimum grain size alignable by k-RATs would increase by a factor of up to 4.8, as shown in Figure~\ref{fig:timescales}, for $\bar{\lambda} = 1.9$\,$\mu$m.}

{The peak column density in emission potentially associated with the Fork is $N_{\textsc{h}_{2}} = 3.1\times 10^{24}\,$cm$^{-2}$, as measured at the methyl formate peak MF2 \citep{favre2011}.  Taking $N_{H}\sim 2.2\times 10^{21}\,$cm$^{2}A_{V}$ \citep{guver2009} and $A_{K}/A_{V} = 0.112$ \citep{rieke1985}, this implies a maximum $K$-band extinction $A_{K}\sim 300$ (the $K$ band is centred on 2.2\,$\mu$m, very similar to our value of $\bar{\lambda}$).  This suggests that strongly directional mid-infrared emission from Source I will drop off very rapidly as it encounters the high-density ambient medium of OMC-1.  The brightness of the submillimetre thermal emission from the Fork which we observe also suggests re-radiation of a significant number of photons.  This strongly suggests that if the polarised emission which we observe in the western arm of the Fork does indeed arise from grains aligned by k-RATs, the polarised emission must arise from the cavity wall of the Source I outflow, where the radiation field of Source I may remain largely unobscured.}

{There are a number of reasons why the k-RAT mechanism might be important to some depth into the western arm of the Fork despite the rapid increase in $\tau_{rad,p}$ with increasing extinction.  Firstly, longer-wavelength photons could maintain a sufficiently short value of $\tau_{rad}$ to permit k-RAT alignment: for our slower value of $\tau_{Lar}$, $\tau_{rad,p}$ as given in equation~\ref{eq:trad_emp} could be increased by a factor $\sim 480$ and remain the shorter timescale at $a = 0.3\,\mu$m, suggesting that k-RATs could potentially remain significant even for values of $\bar{\lambda}$ in the submillimetre regime.  Additionally, our value of $\tau_{Lar}$ is a lower limit, as $B\sim 10\,\mu$G is an upper limit \citep{cortes2020}, unless super-paramagnetic inclusions are resorted to to increase $\chi$.  Similarly, $\tau_{rad,p}$ is an upper limit, as $L = 10^{4}\,$L$_{\odot}$ is a lower limit on the luminosity of Source I \citep{menten1995}.  Finally, our KS tests suggest that the dispersion of $Q$ and $U$ values is systematically lower in the western arm of the Fork than elsewhere in OMC-1; this suggests some difference either in alignment mechanism or in non-thermal velocity dispersion in this region, either or both of which could arise if the emission does indeed arise from the Source I outflow cavity wall rather than the ambient medium.}

{If grains in the western arm of the Fork remain aligned by B-RATs, a large-scale reorientation of the magnetic field must have taken place in the region, away from either the large-scale field direction or the radial field around the BN/KL outflow that we see in the eastern arm.   \citet{cortes2020} find a magnetic Reynolds number $R_{m} = 1.7\times 10^{5}\gg 1$ in OMC-1, indicating that the magnetic field is well-coupled to the gas.  If the field has indeed been reordered in this region, its orientation is suggestive of ordered infall or outflow of material.  If the material were infalling onto the Ridge/Hot Core region, this would be similar to the behaviour seen on much larger size scales in hub-filament systems \citep[e.g.][]{pillai2020}, where magnetic fields are seen to run along filaments of material infalling onto the central hub.  The reordered field representing outflowing material is more difficult to physically motivate, as the most likely cause of such a gas outflow would be the BN/KL explosion, with which our observed polarisation geometry is not consistent.}

This analysis, summarised in Tables~\ref{tab:fork} and~\ref{tab:timescales}, and illustrated in Figure~\ref{fig:timescales}, suggests that {it may be possible for} moderately large grains in the vicinity of extremely luminous sources such as Source I {to} be aligned by k-RATs rather than by B-RATs, {provided that the source remains relatively unobscured.  However, it also suggests that the efficiency of k-RAT alignment will drop off very quickly with distance from Source I due to the high mid-infrared optical depth of the ambient medium of OMC-1.}
We emphasise that our estimates of both $\tau_{rad,p}$ and $\tau_{Lar}$ are highly uncertain.  {We do not have enough information to definitively identify grains in the Fork as being aligned either by k-RATs or by B-RATs.}

\subsection{Main Ridge}
\label{sec:ridge}

\begin{figure}
    \centering
    \includegraphics[width=0.47\textwidth]{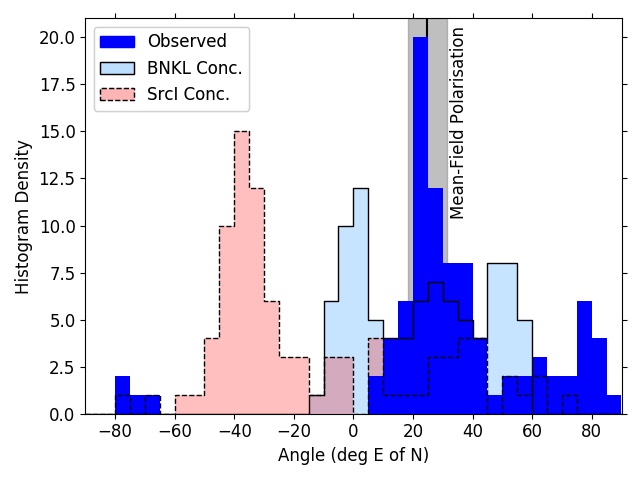}
    \caption{Histogram of polarisation angles in the Ridge (blue), compared with models: polarisation vectors concentric around (1) the centre of the BN/KL explosion (light blue, solid outline), (2) Source I (red, dashed outline).  The polarisation angle associated with the mean 116-degree magnetic field direction is marked.  {Angles are measured on 0.25\arcsec\ (approximately Nyquist-sampled) pixels.}}
    \label{fig:histogram_Ridge}
\end{figure}

\begin{figure*}
    \centering
    \includegraphics[width=0.7\textwidth]{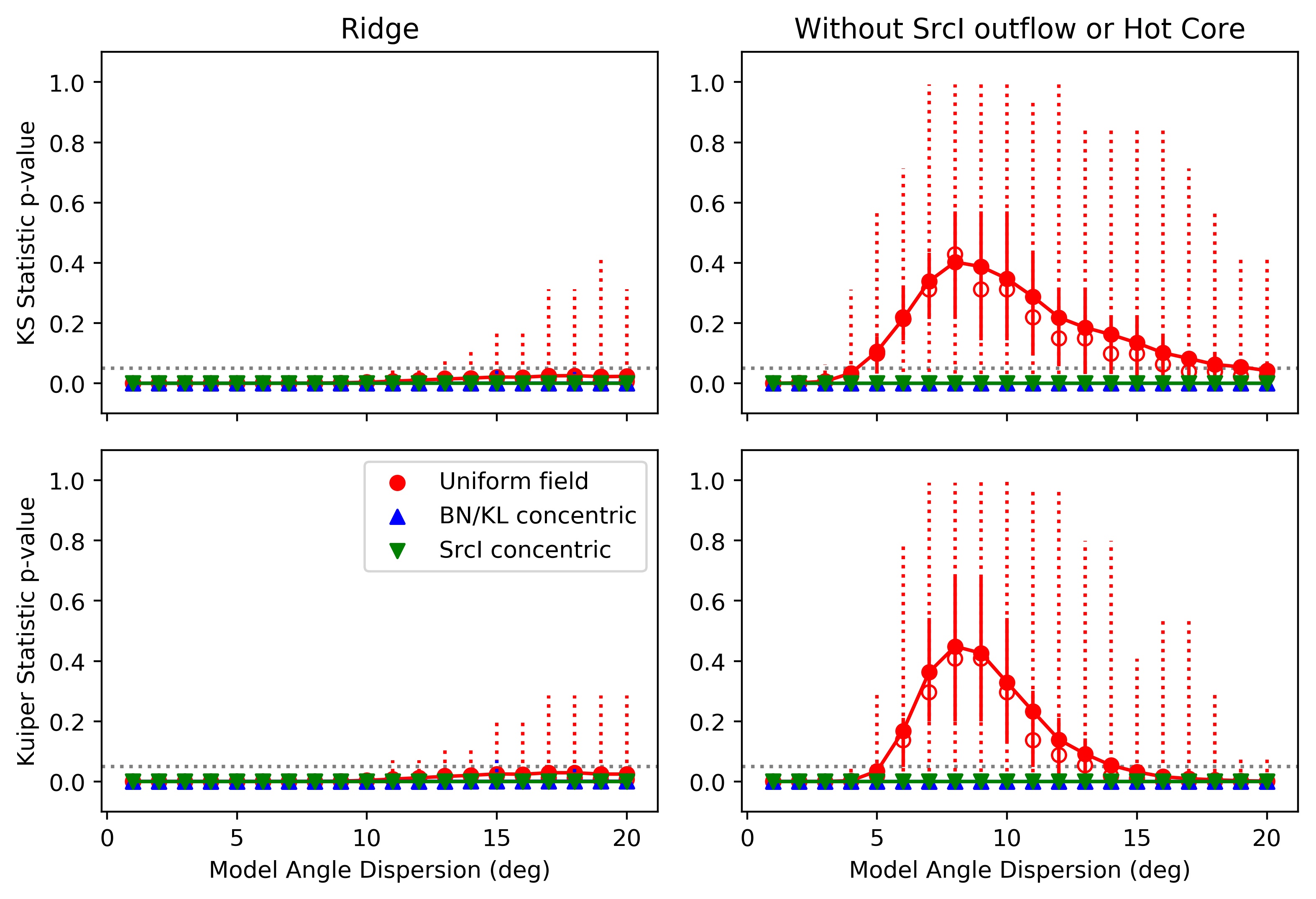}
    \caption{{Results of KS and Kuiper tests comparing observed polarisation angles in the Ridge to a set of Monte Carlo realisations of our three models with specified values of dispersion in polarisation angle.  Description of panels as in Figure~\ref{fig:kstest_AR}.}}
    \label{fig:kstest_Ridge}
\end{figure*}

The main Ridge of OMC-1 (hereafter `the Ridge') is an active site of ongoing star formation, an elongated structure which contains a number of dense cores \citep[e.g.][]{hirota2015}.  Most famous amongst these is the Hot Core \citep{ho1979}, a dense but apparently externally-heated and starless structure \citep{zapata2011} separated from Source I by $\sim 1$\arcsec.

The polarisation pattern in the Ridge is strongly peaked on the {$26^{\circ}$ E of N} polarisation direction {perpendicular to the} large-scale magnetic field, as shown in Figure~\ref{fig:histogram_Ridge}, with deviations in the Hot Core, and on a position NE of Source I and disconnected from the main body of the Ridge, {which we tentatively associate with the Source I outflow.  KS and Kuiper tests, performed as described in Section~\ref{sec:fork_ks}, show that, away from the Hot Core and the Source I outflow, the observed polarisation pattern is consistent ($p>0.05$) with that corresponding to the large-scale magnetic field for angular dispersion values in the range $4^{\circ} \lesssim \sigma_{\theta} \lesssim 20^{\circ}$ (KS test), or  $5^{\circ} \lesssim \sigma_{\theta} \lesssim 16^{\circ}$ (Kuiper test), as shown in Figure~\ref{fig:kstest_Ridge}.  The best agreement with the data is at $\sigma_{\theta} = 8^{\circ}$, matching the equivalent value in the eastern arm of the Fork.}

\begin{figure}
    \centering
    \includegraphics[width=0.47\textwidth]{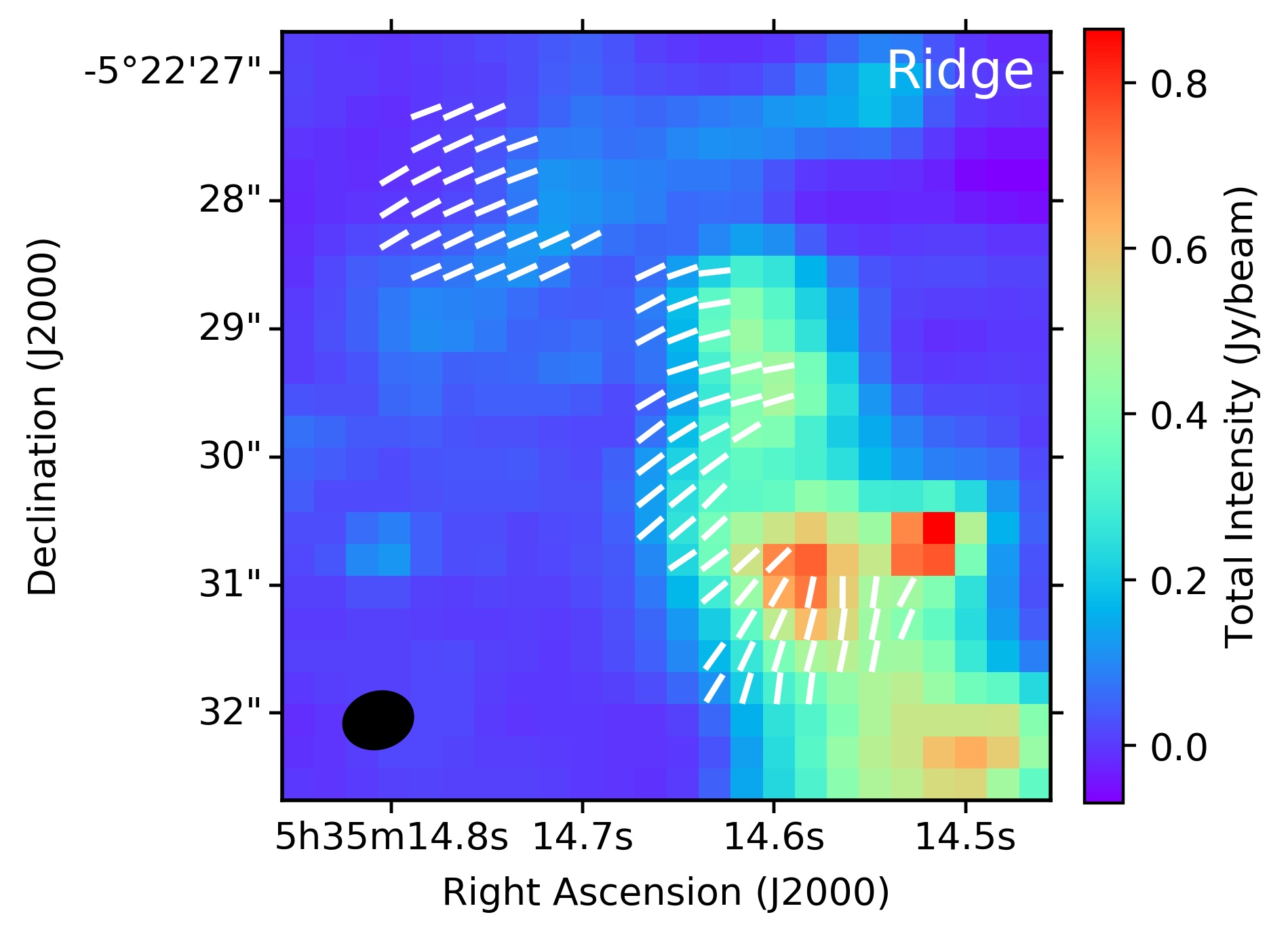}
    \caption{{Magnetic field vectors in the Ridge, obtained by rotating the polarisation vectors by 90$^{\circ}$}, on the assumption that grains are aligned by B-RATs.  We exclude the vectors tentatively associated with emission from the Source I outflow.}
    \label{fig:ridge_bfield}
\end{figure}

{There is no angular dispersion value at which the polarisation pattern in the Ridge is consistent with being concentric around either the centre of the BN/KL explosion or around Source I, suggesting that the grains remain aligned by B-RATs to be perpendicular to the large-scale magnetic field}.  The polarisation vectors in the Ridge, rotated by 90$^{\circ}$ to trace the magnetic field direction, are shown in Figure~\ref{fig:ridge_bfield}.  We exclude from this figure the vectors tentatively associated with the Source I outflow, as discussed in Section~\ref{sec:SrcI_outflow}, below.  We detect little polarised emission in the Ridge south of the Hot Core; particularly, we do not see polarised emission associated with the source SMA~1 \citep{beuther2005}, although the Anomalous Region/Fork borders on this source.  We similarly detect little polarisation on the north-western side of the Ridge.  A possible explanation for this is a lack of a dominant polarisation mechanism in these regions.

Where polarised emission is detected in the Ridge, its direction is consistent with that predicted if the large-scale magnetic field direction persists to the highest-density and smallest-scale structures in OMC-1.  If the magnetic field direction is indeed consistent over orders of magnitude in size scale, it {might suggest} that the field remains dynamically important at the highest densities.  On larger scales in molecular clouds, magnetic fields are consistently found to be perpendicular to filamentary structures where (a) the filament is gravitationally unstable and (b) the magnetic field is, on scales larger than the filament, dynamically important \citep{soler2013,planck2016}.  Although the Ridge is not a filament in the usual sense, it does meet these conditions. 
{However, \citet{cortes2020} find that the energy in the magnetic field is $\sim 3$ orders of magnitude less than the maximum energy in the explosive outflow, and $R_{m} = 1.7\times 10^{5}\gg 1$, indicating that the field moves with the gas, and so argue that the magnetic field in OMC-1 has been reordered as it moves with the outflowing gas to be radial around the BN/KL explosion.  Our results suggest that this reordering may not have taken place in the vicinity of the Ridge.  The major axis of the Ridge is approximately parallel to the BN/KL shock front, and it does not seem to have been disrupted by the BN/KL explosion in the manner of the radial CO streamers that surround OMC-1.  It could, however, be the remnant of some larger structure which has been disrupted or ablated by the passage of the BN/KL shock front.  This suggests that the field might have retained its original orientation in the vicinity of the Ridge not because of its own energetic importance, but rather because of the ability of the gas in the Ridge to which it is coupled to resist disruption by the effects of the BN/KL explosion.}

The polarisation pattern in the Hot Core is broadly similar to that the rest of the Ridge, with some deviation on the south-western side of the core.  This deviation, although not well-resolved, is somewhat suggestive of the pinched field predicted for strongly magnetised dense cores.  A dynamically important magnetic field is broadly expected to support a prestellar core against, and to impose a preferred direction on, gravitational collapse \citep{mouschovias1976}, producing the classical `hourglass' magnetic field indicative of ambipolar-diffusion-mediated gravitational collapse \citep[e.g.][]{fiedler1993}.  However, it is not clear why such an hourglass morphology would only be apparent on one side of the core, {or whether the energetic importance of the magnetic field in the region is sufficiently high to allow such collapse to occur \citep{cortes2020}}.  Higher-resolution polarisation observations are required in order to understand the role of magnetic fields in the evolution of the Hot Core.

We note for completeness that in the south-west side of the Hot Core, the polarisation vectors are also consistent with being elliptical around Source I, with $e\gtrsim 0.9$, as would be expected for k-RAT alignment concentric around Source I in material displaced along the line of sight with respect to Source I.  This is the only region in the Ridge where polarisation vectors are consistent with k-RAT alignment.  However, as argued below, it appears unlikely that k-RATs can dominate over B-RATs in the Ridge, and there is no clear reason for the Hot Core to be the exception to this.

\subsubsection{Source I outflow?}
\label{sec:SrcI_outflow}

Polarisation is detected at a position north-east of Source I, and disconnected from the Ridge.  This region, labelled as `Source I outflow?' in Figure~\ref{fig:PI}, has polarisation vectors approximately perpendicular to those in both the Ridge and Source I, and thus are inconsistent both with the {$26^{\circ}$ E of N polarisation direction perpendicular to the} large-scale field direction and with being concentric around the BN/KL explosion, as can be seen in Figure~\ref{fig:compare_Ridge}.  These vectors have orientations qualitatively similar to the SiO polarisation vectors detected by \citet{hirota2020} in the north-eastern lobe of the Source I outflow, perhaps suggesting that this polarised emission arises from dust {in the outflow cavity walls, as is hypothesised for the Fork \citep{wright2017}, or} entrained by the outflow.  We note, however, that the size scale of the SiO measurements is quite different to our observations (\citet{hirota2020} observed $\sim 1$\arcsec\ around Source I), and so assigning this emission to the Source I outflow is speculative.

If the dust grains in the outflow are aligned by B-RATs, they could be tracing a helical magnetic field structure \citep{hirota2020}.  However, the vector orientations are also qualitatively similar to being concentric around Source I, as shown in Figure~\ref{fig:compare_Ridge}.  This might suggest that grains in this region could instead be aligned by k-RATs, as we hypothesise in the Anomalous Region/Fork.  It seems plausible that the extreme conditions apparently giving rise to k-RATs in the Fork to the south-east of Source I might also be expected to arise in the north-western outflow cone; however, we do not have sufficient evidence to conclusively determine the grain alignment mechanism in this region.

\begin{figure*}
    \centering
    \includegraphics[width=\textwidth]{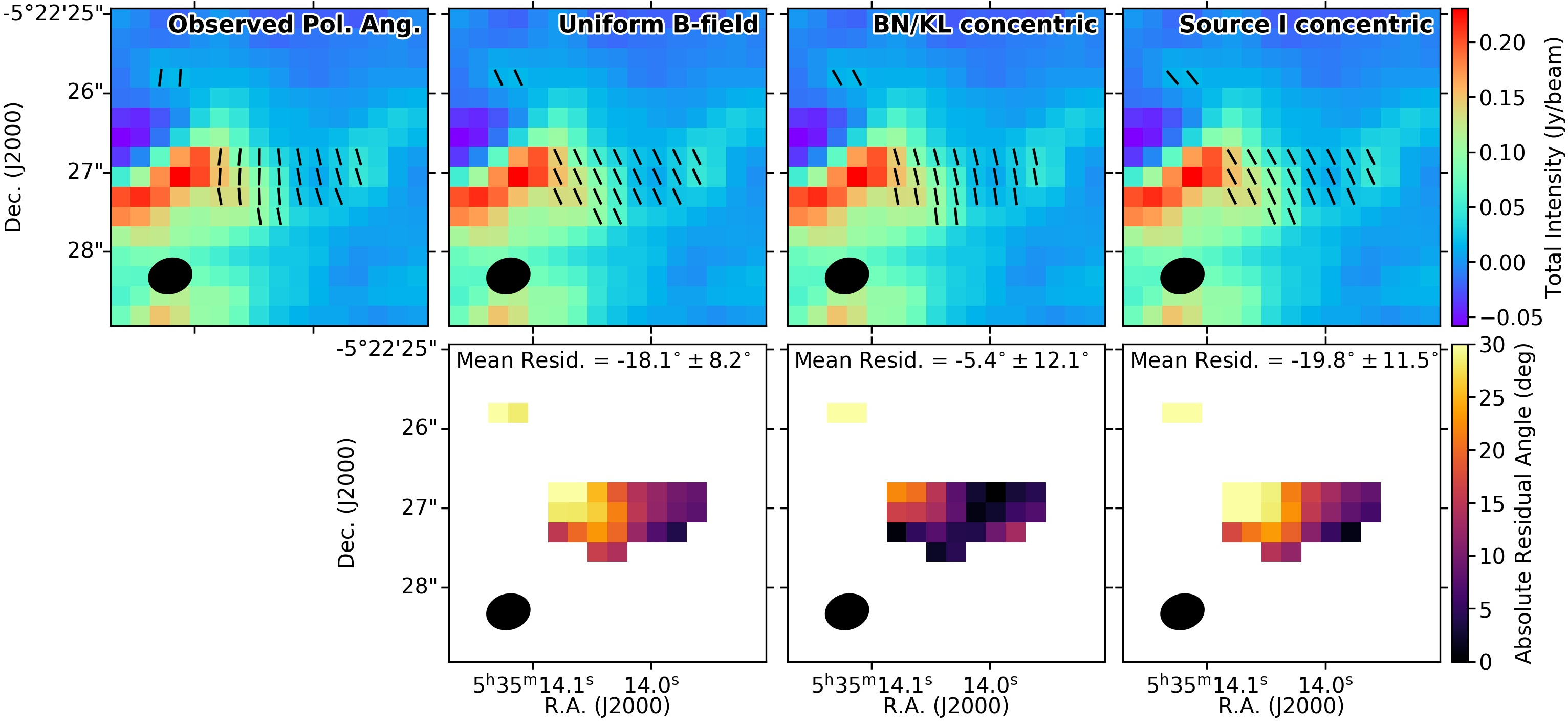}
    \caption{Comparison of models in MF4/MF5.
    Top row shows {observed and} model polarisation geometries, {plotted on Stokes $I$ emission}, bottom row shows absolute difference in angle between data and models.  {Far left: Observed polarisation vectors.  Centre left:} polarisation vectors aligned {$26^{\circ}$ E of N,} perpendicular to the the large-scale magnetic field direction (hypothesised alignment mechanism: B-RATs).  Centre {right}: polarisation vectors concentric around the BN/KL explosion centre (hypothesised alignment mechanism: {B-RATs/v-MATs}).  {Far right}: polarisation vectors concentric around Source I (hypothesised alignment mechanism: k-RATs).  {In the bottom row, the colour table saturates at a difference in angle of $30^{\circ}$, to emphasise the differences between the models.}  All maps are shown on 0.25\arcsec\ (approximately Nyquist-sampled) pixels.  The synthesised beam size is shown in the lower left-hand corner of each plot.}
    \label{fig:compare_NWClump}
\end{figure*}

\begin{figure}
    \centering
    \includegraphics[width=0.47\textwidth]{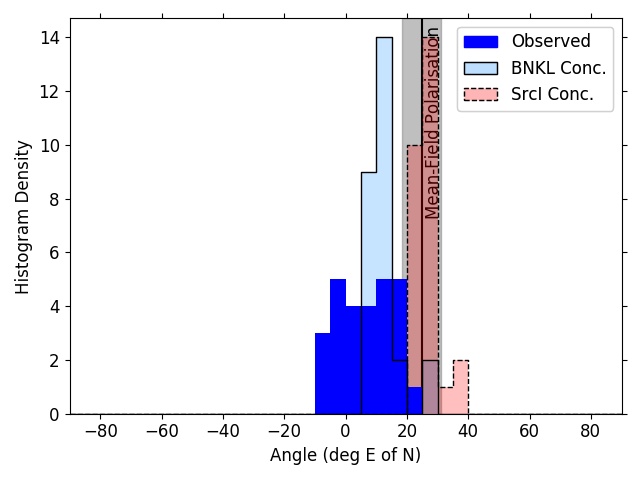}
    \caption{Histogram of polarisation angles in the MF4/MF5 (blue), compared with models: polarisation vectors concentric around (1) the centre of the BN/KL explosion (light blue, solid outline), (2) Source I (red, dashed outline).  The polarisation angle associated with the mean 116-degree magnetic field direction is marked.  {Angles are measured on 0.25\arcsec\ (approximately Nyquist-sampled) pixels}}
    \label{fig:histogram_NWClump}
\end{figure}

\begin{figure}
    \centering
    \includegraphics[width=0.47\textwidth]{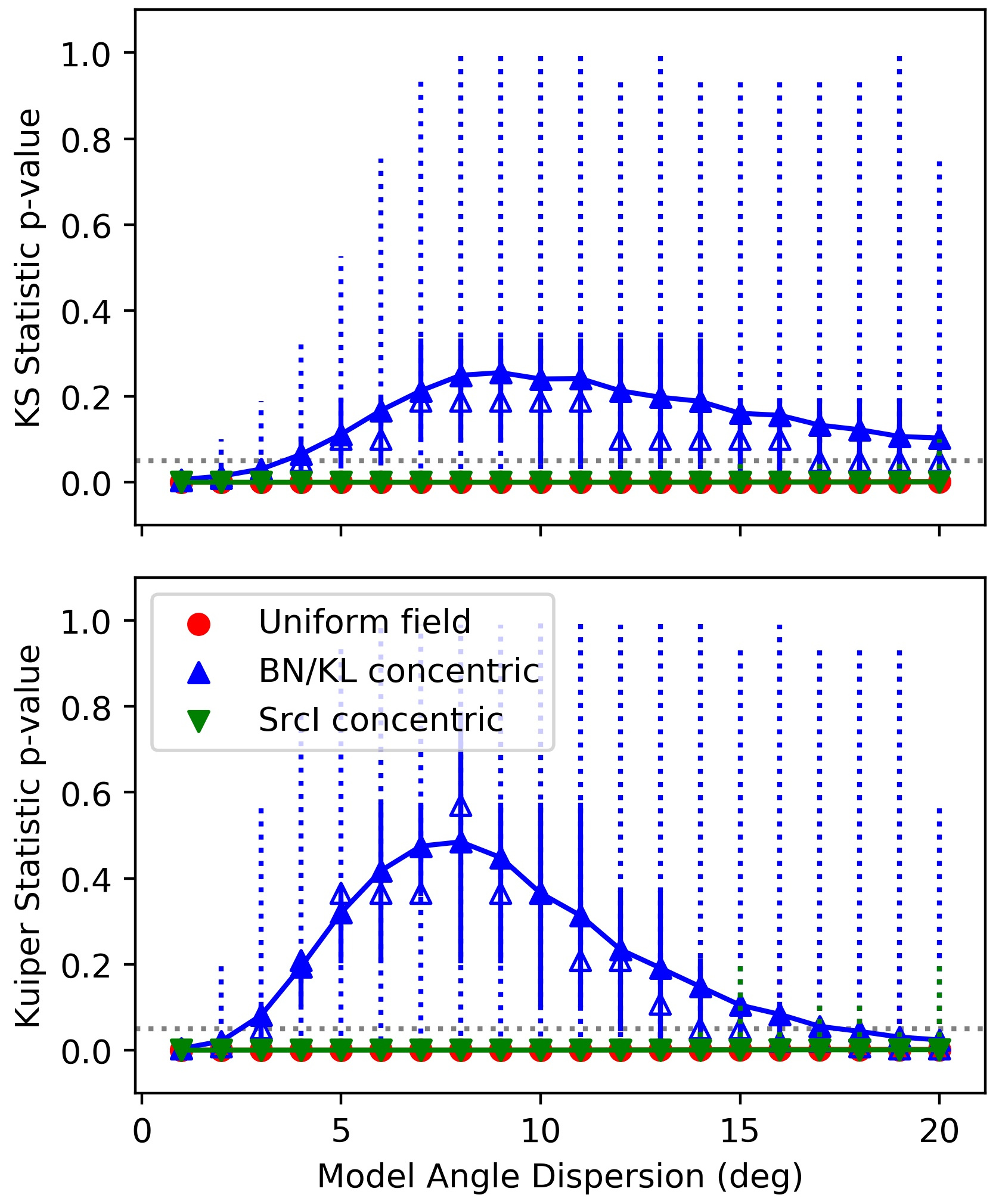}
    \caption{{Results of KS and Kuiper tests comparing observed polarisation angles in MF4/MF5 to a set of Monte Carlo realisations of our three models with specified values of dispersion in polarisation angle.  Description of panels as in Figure~\ref{fig:kstest_AR}.}}
    \label{fig:kstest_NWClump}
\end{figure}

\begin{figure}
    \centering
    \includegraphics[width=0.47\textwidth]{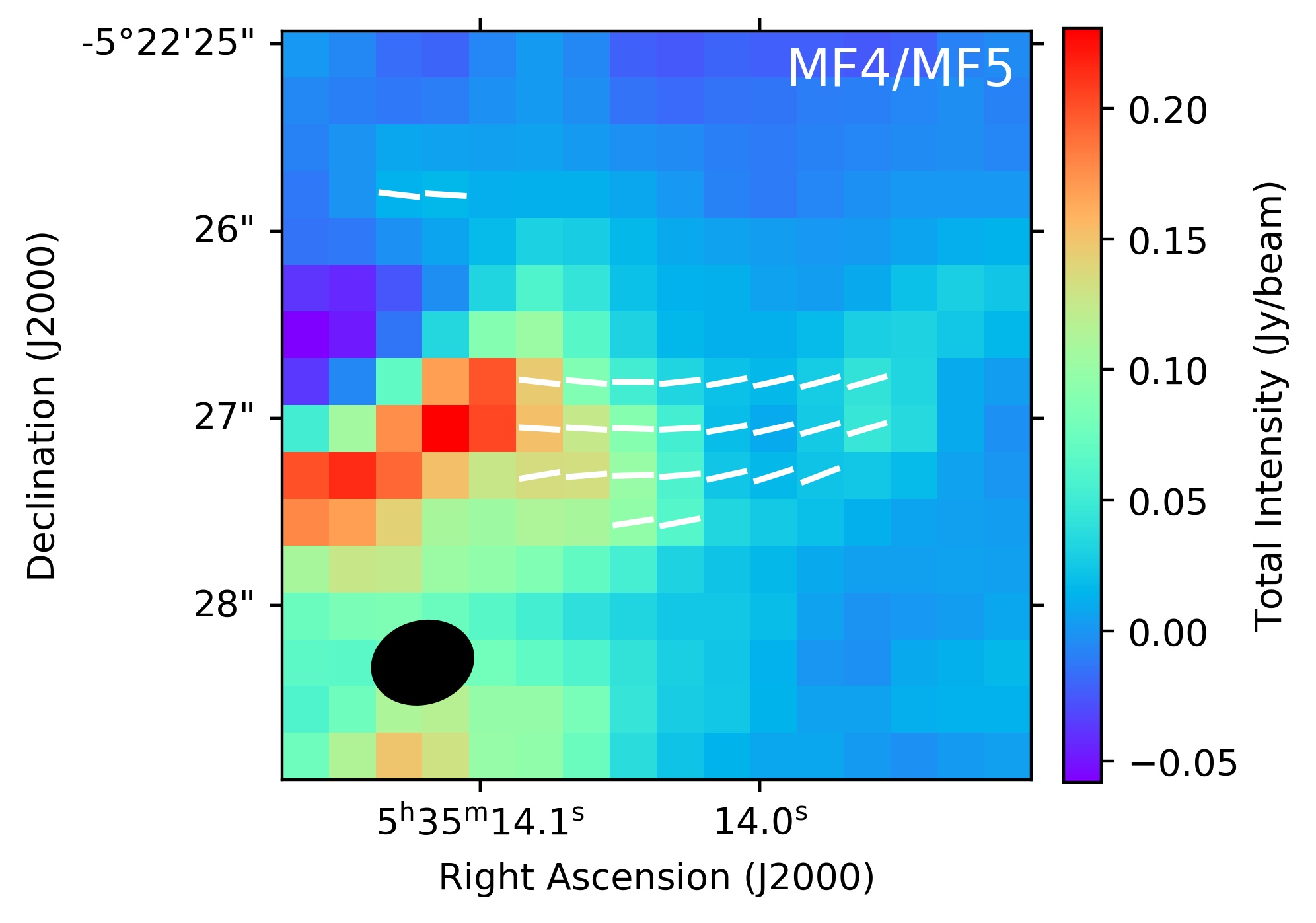}
    \caption{{Magnetic field vectors in MF4/MF5, obtained by rotating the polarisation vectors by 90$^{\circ}$}.}
    \label{fig:NWClump_bfield}
\end{figure}

\subsubsection{B-RAT versus k-RAT alignment in the Ridge}

With the exception of the handful of vectors tentatively associated with the Source I outflow, the polarisation pattern in the Ridge is inconsistent with being induced by k-RATs driven by Source I.  Given that the Ridge is at a similar or smaller distance to Source I than is the Anomalous Region/Fork, this {raises} the question of how its grains have apparently retained their original alignment with respect to the magnetic field.

The requirement for k-RATs to dominate over B-RATs is $\tau_{rad, p}<\tau_{Lar}$.  The material of the Ridge has {a higher column density ($N_{\textsc{h}_{2}} = 5.4\times 10^{24}$\,cm$^{-2}$; \citealt{favre2011}) than its surroundings, corresponding to a peak $K$-band extinction $A_{K}\sim 500$}.  For much of the Ridge, the emission from Source I will be {further} obscured by its disc.  {The effect of this obscuration will be to increase $\bar{\lambda}$ and decrease $\gamma$, thereby increasing $\tau_{rad, p}$, as discussed in Section~\ref{sec:fork_discuss}}.  We note also that the polarised emission which we see in the Ridge mostly arises from its eastern side, away from Source I.  {The hypothesised k-RAT alignment in the Fork appears to depend on unobscured emission from Source I driving alignment in the outflow cavity walls.}  This effect is less likely to apply in the Ridge, although the northern part of the Ridge could be impacted on by the Source I outflow, depending on the relative orientations of the Ridge and the outflow.

We also expect $\tau_{Lar}$ to decrease in high-density material, as magnetic field strength $B$ is expected to scale with density such that $B\propto n^{0.5}$ or $B\propto n^{0.66}$ \citep[e.g.][]{crutcher2012}.

While these two effects are difficult to quantify, the polarisation geometry which we observe on the eastern side of the Ridge suggests that between them they are sufficient to result in $\tau_{rad, p}>\tau_{Lar}$, allowing B-RATs to dominate over k-RATs, despite the proximity of Source I.  {This again suggests that k-RAT alignment, if present in ISM material, is restricted to the immediate vicinity of (proto)stars, and in the vast majority of ISM environments, grains will remain aligned relative to the magnetic field.}

We note that the increase in density will also decrease $\tau_{gas}$ as shown in equation~(\ref{eq:tgas}), although this will be mitigated by a decrease in $T_{gas}$ as the radiation field of Source I is increasingly obscured (cf. equation~\ref{eq:Td}).  Nonetheless, as discussed in Section~\ref{sec:fork_tgas}, a density increase of {several orders of magnitude would be required} to make $\tau_{gas}<\tau_{Lar}$ hold for physically plausible grain sizes, particularly if $\tau_{Lar}$ is itself shortened in the Ridge.

\begin{figure*}
    \centering
    \includegraphics[width=\textwidth]{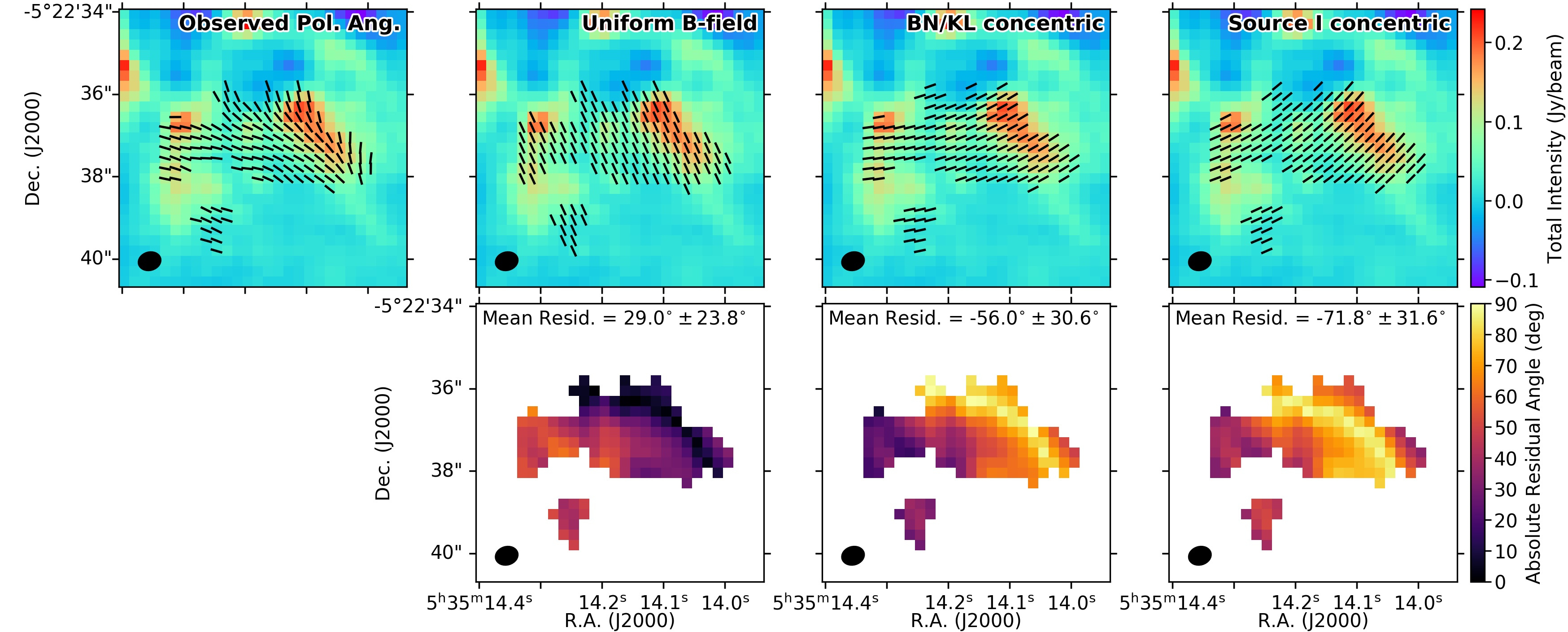}
    \caption{Comparison of models in MF1.  Top row shows {observed and} model polarisation geometries, {plotted on Stokes $I$ emission}, bottom row shows absolute difference in angle between data and models.  {Far left: Observed polarisation vectors.  Centre left:} polarisation vectors aligned {$26^{\circ}$ E of N,} perpendicular to the the large-scale magnetic field direction (hypothesised alignment mechanism: B-RATs).  Centre {right}: polarisation vectors concentric around the BN/KL explosion centre (hypothesised alignment mechanism: v-MATs).  {Far right}: polarisation vectors concentric around Source I (hypothesised alignment mechanism: k-RATs).  {All maps are shown on 0.25\arcsec\ (approximately Nyquist-sampled) pixels.  The synthesised beam size is shown in the lower left-hand corner of each plot.}}
    \label{fig:compare_MF1}
\end{figure*}

\begin{figure}
    \centering
    \includegraphics[width=0.47\textwidth]{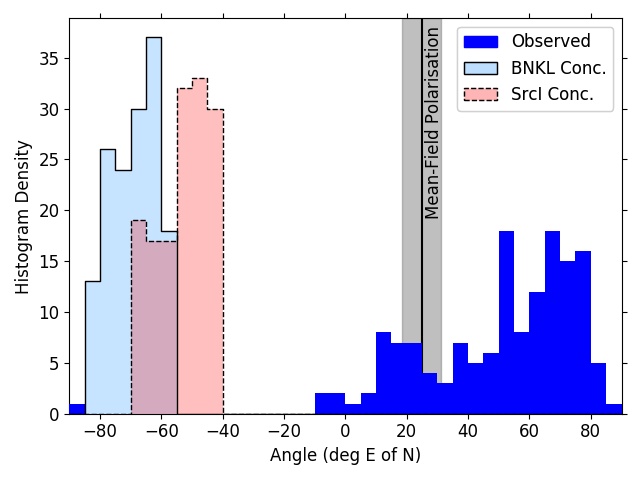}
    \caption{Histogram of polarisation angles in MF1 (blue), compared with models: polarisation vectors concentric around (1) the centre of the BN/KL explosion (light blue, solid outline), (2) Source I (red, dashed outline).  The polarisation angle associated with the mean 116-degree magnetic field direction is marked.  Angles are measured on 0.25\arcsec\ (approximately Nyquist-sampled) pixels.}
    \label{fig:histogram_MF1}
\end{figure}

\subsection{MF4/MF5}

Polarisation vectors in MF4 and MF5 (\citealt{favre2011}; collectively known as the Northwest Clump) are {qualitatively} similar both to the pattern predicted for alignment perpendicular to the large-scale field direction, and to that for being concentric around the BN/KL outflow centre or Source I.  The two clumps are at a similar distance to the BN/KL explosion centre as is the Ridge, and have complex substructure, with each consisting of three distinct velocity components \citep{pagani2017}.  We detect only 6 independent beams over MF4/MF5.

As shown in Figures~\ref{fig:compare_NWClump} and \ref{fig:histogram_NWClump}, the polarisation pattern in MF4/MF5 is more consistent with that expected for concentric polarisation around BN/KL than with the mean field direction or with being concentric around Source I.   {This is confirmed by KS and Kuiper tests, performed as described in Section~\ref{sec:fork_ks}, which show that only the BN/KL-concentric model can be made consistent ($p > 0.05$) with the data, as shown in Figure~\ref{fig:kstest_NWClump}.  A wide range of angular dispersion values, $\sigma_{\theta} \gtrsim 4.5^{\circ}$ (KS test) or $2.5^{\circ} \lesssim \sigma_{\theta} \lesssim 17.5^{\circ}$ (Kuiper test), produce patterns consistent with the small sample of observed vectors, but the best agreement is found at $\sigma_{\theta} \sim 8^{\circ}$, matching the equivalent values in the eastern arm of the Fork and the Ridge.  Our results thus support the interpretation of \citet{cortes2020} that in this region, the magnetic field has been realigned by the effects of the BN/KL explosion.  The magnetic field implied by our polarisation observations is shown in Figure~\ref{fig:NWClump_bfield}.} 

\subsection{Compact Ridge/MF1}

The Compact Ridge (also known as MF1) is a $\sim 4.3\,$\msun\ clump \citep{favre2011}, {found by \citet{pagani2017} to have extremely narrow ($\sim 1\,$km\,s$^{-1}$) linewidths. \citet{pagani2017} thus suggest that MF1 appears not to have yet to have been affected} by the BN/KL explosion, and so place it at least 10\,000\,AU, and likely $\sim 20\,000$\,AU, {either in front of or behind} the explosion centre along the line of sight.  It is thus unlikely to be physically associated with the other dense clumps which we observe.

\begin{figure}
    \centering
    \includegraphics[width=0.47\textwidth]{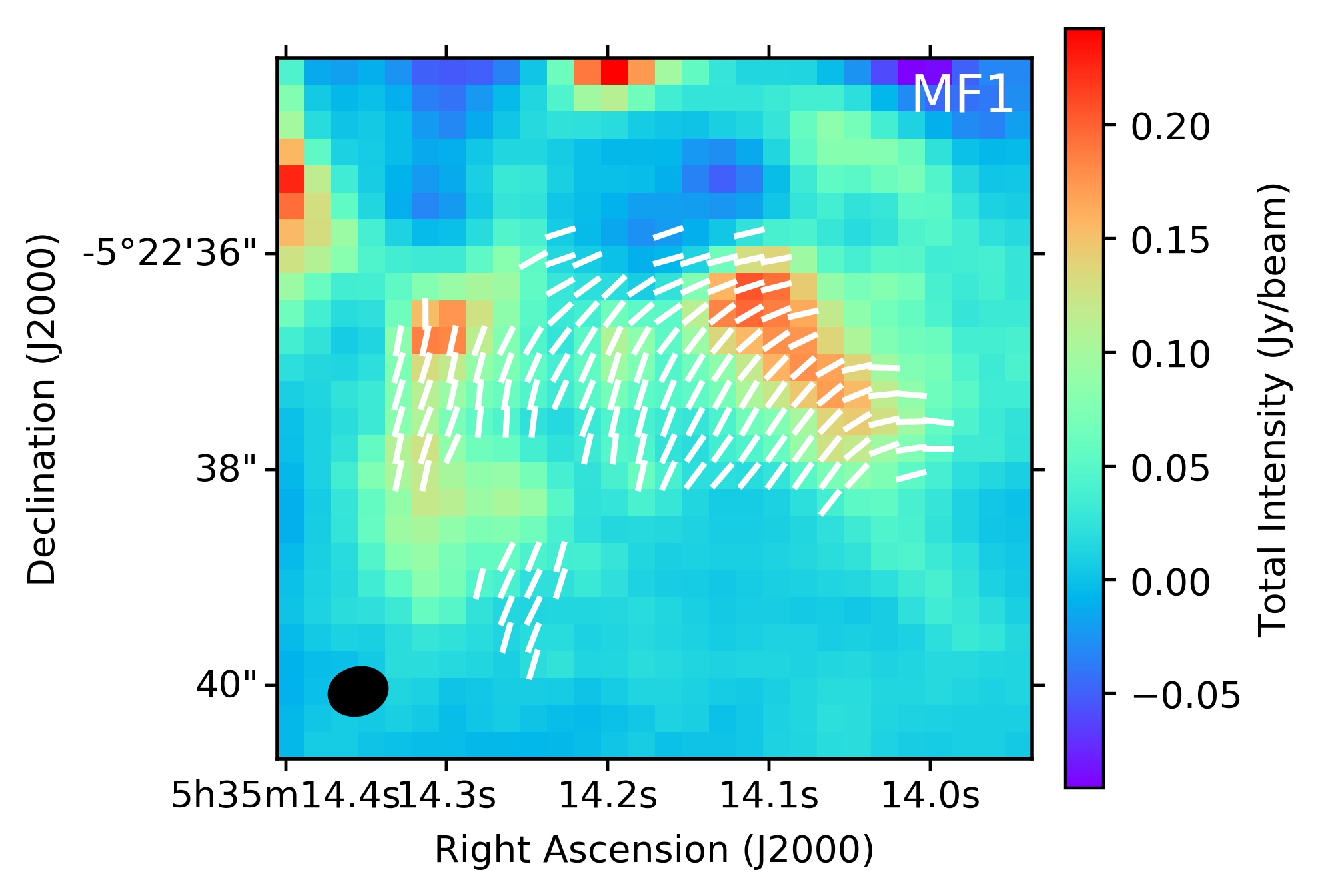}
    \caption{{Magnetic field vectors in MF1, obtained by rotating the polarisation vectors by 90$^{\circ}$}.}
    \label{fig:mf1_bfield}
\end{figure}

The polarisation pattern in MF1 is inconsistent with being concentric around either BN/KL or Source I, and broadly similar to the polarisation pattern expected for grains aligned perpendicular to the large-scale 116-degree field, as shown in Figures~\ref{fig:compare_MF1} and \ref{fig:histogram_MF1}.  This result is in keeping with the hypothesis that the region is at a significant distance from the other clumps considered here.  We can with some reliability in MF1 expect grains to remain aligned by B-RATs, and so we show the polarisation vectors, rotated by 90$^{\circ}$ to trace the magnetic field direction, in Figure~\ref{fig:mf1_bfield}.  {However, KS and Kuiper tests, performed as described in Section~\ref{sec:fork_ks}, show that none of our proposed simple models can be made consistent with the observed polarisation pattern.  We do not show the results of these tests in a figure, as in every case $p\ll 0.05$.}  There is significant ordered variation in the magnetic field direction across MF1, and the implied mean and median magnetic field direction values ($147^{\circ}$ and $149^{\circ}$, respectively) are similar to, but do not match, the average large-scale field direction.  {The field lines appear to be bowing away from the mean field direction in the periphery of MF1, turning towards the south (this behaviour is more apparent on the eastern side of the region). A possible interpretation of this behaviour is that the field in MF1 is merging into the large-scale `hourglass' field morphology observed on larger scales in OMC-1 \citep{rao1998,houde2004,pattle2017}.  \citet{cortes2020} also note a transition toward field morphologies matching the large-scale hourglass in the south of OMC-1.}

\section{Summary} \label{sec:summary}

We have presented ALMA Band 7 {(881\,$\mu$m) continuum} polarisation observations of the centre of the OMC-1 region of the Orion Molecular Cloud.

We divided OMC-1 into five regions: Source I (a massive outflow-driving protostar), the Anomalous Region/Fork, the Main Ridge, MF4/MF5, and the Compact Ridge/MF1.  Our key findings are as follows:
\begin{enumerate}
    \item In Source I, we found {an unresolved} polarisation geometry parallel to the minor axis of the Source I disc, consistent with polarisation arising from dust self-scattering.  The Source I disc is optically thick and viewed almost edge-on, supporting this interpretation; {however, the predicted polarisation pattern is degenerate with that for a polarisation pattern concentric around the centre of the BN/KL explosion}.
    \item {In the eastern arm of the Anomalous Region/Fork, we found a polarisation geometry consistent with being concentric around the centre of the BN/KL explosion, consistent with the magnetic field having been reordered to be radial around the BN/KL explosion, as posited by \citet{tang2010} and \citet{cortes2020}.}
    \item {In the western arm of the Fork,} a region in which emission may arise from the Source I outflow cavity walls, we found a polarisation geometry consistent {with being concentric around Source I.}  We compared the mechanical alignment timescale $\tau_{mech}$ to the Larmor timescale $\tau_{Lar}$ in the Anomalous Region/Fork, finding $\tau_{mech}\gg\tau_{Lar}$, indicating that grains are unlikely to be aligned by subsonic mechanical alignment torques (v-MATs) induced by the passage of shocks associated with the BN/KL explosion or associated with the Source I outflow.  We compared the radiative precession timescale $\tau_{rad,p}$ for {unobscured} emission from Source I to the Larmor timescale $\tau_{Lar}$ in the Anomalous Region/Fork.  {While our estimates of both timescales are highly uncertain,} we find $\tau_{rad,p}<\tau_{Lar}$ for moderately large grains {($>0.005-0.1\,\mu$m)}, {suggesting} that grains in this region {may} be aligned by radiative torques to precess around the radiation anisotropy gradient (k-RATs), i.e. to be perpendicular to the gradient of intensity from Source I.  {This hypothesis strongly} favours the interpretation of emission in the region as arising from the Source I outflow cavity walls, as Source I must remain relatively unobscured for k-RATs to dominate in this manner.  {Alternatively, the grains may continue to precess around the magnetic field direction (B-RATs), and so could trace infall of material onto the Main Ridge/Hot Core region.}
    \item In the Main Ridge, we found a {polarisation geometry consistent with that of the large-scale magnetic field in the region}, and so determined that grains are aligned perpendicular to the magnetic field (B-RAT alignment).  The {consistency with the large-scale magnetic field, despite recent findings that the magnetic field is energetically subdominant compared to the BN/KL outflow, may result from the field being well-coupled to gas in the Ridge that has not been disrupted by the BN/KL explosion.}  We identified an area of polarised emission north-east of Source I possibly arising from the Source I outflow.  Grains in this region could trace a helical magnetic field in the outflow or be aligned by k-RATs.
    \item In MF4/MF5, we found {a polarisation geometry consistent with being concentric around the BN/KL explosion centre, and so tracing a magnetic field that has been reordered to be radial around the BN/KL explosion.}
    \item In the Compact Ridge/MF1, likely located sufficiently far from the BN/KL explosion and Source I to remain uninfluenced by their effects, {we found} a polarisation geometry similar to, but showing ordered deviation from, {being perpendicular to} the large-scale magnetic field direction. { The field may here be merging into the large-scale `hourglass' field identified in single-dish observations.}
\end{enumerate}
Our observation of grains which {may} be aligned by k-RATs rather than by B-RATs in the vicinity of Source I demonstrates {the care which must be taken in the interpretation of polarisation observations in extreme environments in the interstellar medium}.
{The complexity of OMC-1, and the similarity between the polarisation geometries predicted by our simple models, makes definitively identifying k-RAT alignment difficult.  We cannot rule out either k-RAT or B-RAT alignment as having produced the observed polarisation geometry.  In order to confirm the existence of k-RAT alignment in the vicinity  of massive protostars, we will need to observe the polarisation geometry around similar sources in other less complex regions.}

\section*{Acknowledgements}

{We would like to thank the anonymous referee for a helpful report which significantly improved the content of this paper.}  This paper makes use of the following ALMA data: ADS/JAO.ALMA\#2018.1.01162.S. ALMA is a partnership of ESO (representing its member states), NSF (USA) and NINS (Japan), together with NRC (Canada), MOST and ASIAA (Taiwan), and KASI (Republic of Korea), in cooperation with the Republic of Chile. The Joint ALMA Observatory is operated by ESO, AUI/NRAO and NAOJ.  K.P. and S.P.L. acknowledge support from the Ministry of Science and Technology (MOST), Taiwan, under grant numbers 106-2119-M-007-021-MY3 {(K.P., S.P.L.) and 109-2112-M-007-010-MY3 (S.P.L.)}.
This research was supported in part at the SOFIA Science Center, which is operated by the Universities Space Research Association under contract NNA17BF53C with the National Aeronautics and Space Administration.  {T.H. acknowledges support by the National Research Foundation of Korea (NRF) grants funded by the Korean government (MSIT) (2019R1A2C1087045).}
This research made use of Astropy, a community-developed core Python package for Astronomy \citep{astropy2013, astropy2018}, the CASA software suite \citep{mcmullin2007}, Starlink software \citep{currie2014}, currently supported by the East Asian Observatory, and the NASA Astrophysics Data System. 

\section*{Data Availability}

The data used in this paper are available in the ALMA Science Archive, under project code 2018.1.01162.S.

\bibliographystyle{mnras}

\bsp	
\label{lastpage}
\end{document}